\documentclass[10pt,english,a4paper]{article}
\usepackage[utf8]{inputenc}
\usepackage[english]{babel}
\usepackage{amsmath}
\usepackage{amsfonts}
\usepackage{amssymb}
\usepackage{graphicx}
\usepackage{wrapfig}
\usepackage{titling}
\usepackage{color}
\usepackage{float}
\usepackage{titlesec}
\usepackage{titling}
\usepackage{abstract}
\usepackage[square]{natbib}
\usepackage{stmaryrd}
\usepackage{glossaries}
\usepackage{setspace}
\usepackage{listings}
\usepackage{pxfonts}

\usepackage{lmodern,dsfont}

\usepackage{subscript}
\usepackage[top=3.25cm, bottom=3.5cm, left=3.5cm, right=3.5cm]{geometry}
\usepackage[center]{caption}
\usepackage[toc, page]{appendix}
\usepackage[english, pdfborder={0 0 0}]{hyperref}

\hypersetup{
  pdfproducer={}
}

\usepackage[version=0.96]{pgf}
\usepackage{tikz}
\usetikzlibrary{arrows,shapes,snakes,automata,backgrounds,petri}

\usepackage{color}
\definecolor{gray}{rgb}{0.4,0.4,0.4}
\definecolor{darkblue}{rgb}{0.0,0.0,0.6}
\definecolor{cyan}{rgb}{0.0,0.6,0.6}

\lstdefinelanguage{maude}
{
    alsoletter={\:},
    morecomment=[l]{---},
    morecomment=[l]{***},
    keywords={in, load, pr, protecting, sort, sorts, subsort, subsorts, including, class, msg, msgs, endfm, fmod, is, mod, endm, omod, endom},
    keywords=[2]{eq,  mb, ceq, if, rl, crl, else, then, fi},
    keywords=[3]{ctor, assoc, comm, gather, id\:},
    keywords=[4]{op, ops, var, vars},
    breaklines=true,
    frame=lrtb,
    captionpos=b,
    numbers=left,
    stepnumber=2,
    aboveskip=15pt
}

\lstdefinestyle{maude}{
    language=maude,
    keywordstyle=\bfseries\color{rojoReservado},
    keywordstyle=[2]\bfseries\color{verdeReservado},
    keywordstyle=[3]\bfseries\color{turquesaReservado},
    keywordstyle=[4]\bfseries\color{moradoReservado},
}

\begin{document}
\title{Project 2\\\textbf{Converting Reconfigurable Petri Nets to Maude}}
\author{Alexander Schulz\\\href{mailto:alexander.schulz1@haw-hamburg.de}{alexander.schulz1@haw-hamburg.de}\\ \\ Supervised by\\ \\ Prof. Dr. Padberg\\\href{mailto:julia.padberg@haw-hamburg.de}{julia.padberg@haw-hamburg.de}}
\date{\today}
\maketitle

\tableofcontents

\newpage

\begin{abstract}
\noindent Model checking is an important aim of the theoretical computer science. It enables the verification of a model with a set of properties such as liveness, deadlock or safety. One of the typical modelling techniques are Petri nets they are well understood and can be used for a model checking. Reconfigurable Petri nets are based on a Petri nets with a set of rules. These rules can be used dynamically to change the net.\\
Missing is the possibility to verify a reconfigurable net and properties such as deadlocks or liveness. This paper introduces a conversion from reconfigurable Petri net to Maude, that allows us to fill the gap. It presents a net transformation approach which is based on Maude's equation- and rewrite logic as well as the LTLR model checker.
\end{abstract}

\section{Motivation}
The first approach to convert reconfigurable Petri net to Maude (see \citep{project1}) is designed as extension for \textit{ReConNet}. It uses the implementation of \textit{ReConNet} to get all possible matches for a set of rules and a given Petri net. This approach results in a dependency of the current net state and the algorithm of \textit{ReConNet}.\\
Hence, that the model-checking process has only one-step application of a rule may be wrong if a rule is used twice, because an error may occur after the second usage.\\
The new approach is based on the algebraic structure of reconfigurable Petri nets. The main task was to create a structure which can be read similar to the mathematical notation of a reconfigurable Petri net. It includes the possibility to simulate the net. This implies a solution for a transition that defines the activation and firing. Maude has been chosen as the appropriate language to implement this definition.\\
Moreover, the rules need to be implemented within the new structure. In contrast to the first approach it should be able to detect a match itself.\\
The following sections contain an overview of all relevant parts of this approach. The first section gives a short overview of the background for this work. The next section focuses the new modules, which contain the data-types for the resulting Maude specification of a converted reconfigurable Petri net. Finally, an evaluation shows the performance of the current implementation based on a test net.

\newpage
\section{Background}
First we introduce reconfigurable Petri nets. They extend Petri nets with a set of rules, that can modify the net at runtime. Moreover, Maude is introduced since it is the result the conversion aims at. Finally, a short survey of related works is presented.

\subsection{Reconfigurable Petri nets}
\begin{figure}[H]
\centering
\includegraphics[keepaspectratio=true, width=\linewidth]{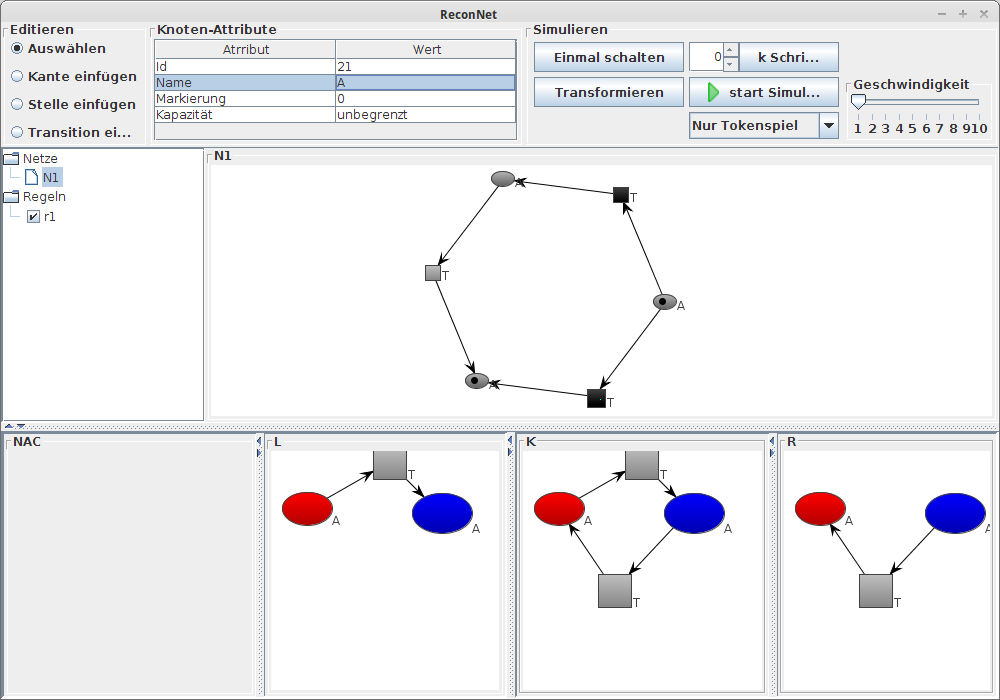}
\caption{ReConNet - graphical editor for reconfigurable Petri nets}
\label{fig:reConNet}
\end{figure}

One of the most important models for concurrent systems and some software engineering parts are the Petri nets, based on Carl Adam Petri's dissertation \citep{petri1962kommunikation}.\\
\noindent A marked Petri net can be formally described as a tuple $N = (P,T,pre,post,M_{0})$ where $P$ is a set of places and $T$ is a set of transitions. $pre$ is used for all $pre$-conditions of transitions, which describes how many token are required for firing. On the other hand, $post$ holds all information of the post-conditions for all transitions. Finally, $M_{0}$ shows all initial tokens on the places for this net $N$ \citep{meseguer1990petri, petriNetGrundlagen}.\\
\noindent Further, based on a Petri net are reconfigurable Petri nets important because they can modify themselves with a set of rules \citep{ehrig2007independence, prange2008transformations, kahloul2010modeling}. A reconfigurable Petri net can be describe a tuple of a reconfigurable Petri net $RN = (N,$ $\cal R)$. This definition uses the Petri net tuple and a set $\cal R$ of rules, which are given by rule $r = (L \leftarrow K \rightarrow R)$ \citep{padberg2012abstract, ehrig2009cospan}. $L$ is the left-hand side ($LHS$), which needs a morphism to be mapped to a net $N$. $K$ is an interface between $L$ and $R$. $R$ is the part which is inserted into the original net. To realise this replacement a matching algorithm needs to be defined that finds $L$ within the source net $N$. This match includes a mapping between the elements in the Petri net and the left side of the rule ($L$). Basically, this algorithm finds the same structure (form $L$) within the Petri nets \citep{Blumreiter13}.\\
Reconfigurable Petri nets are also comprises capacities and labels for transitions/places \citep{padberg2012abstract}. A limitation for place is realised via a capacity, that contains a value which describes how much token can be stored on a place. The function $\text{\textit{cap}} : P \rightarrow \mathbb{N}_{+}^{w}$ assigns for each place a natural number as capacity. Further, two label function (\textit{pname} and \textit{tname}) refer for each place or transition a name from a name space ($\text{\textit{pname}} : P \rightarrow A_{P}$ and $\text{\textit{tname}} : R \rightarrow A_{T}$).\\
\textit{ReConNet} is shown in \autoref{fig:reConNet} with an example net $N_{1}$ and rule $\cal R$$_{1}$. The configurations such as node names or markings as well as the control elements for firing and transformation are presented in the upper region of the graphical interface. A graphical illustration of reconfigurable Petri nets is in the remaining interface. First, a net which models a cycle is composed of three places and transitions as well as one token. Wider, a rule which changes the arc direction of a transition is displayed under the net. In both editors are activated transitions marked as black transitions instead of grey normal transitions. Colours for places in rules are used to ensure the common bond.

\subsection{Maude}
Maude has been developed at the Stanford Research Institute International (SRI International) for the last two decades. The equation and rewriting logic, which supports a powerful algebraic language, is used as a base \citep{escobar2009variant,clavel2002maude}. Based on these two kinds of logics Maude models a concurrent state systems that used for semantic analysis such as deadlock discovery via LTL-model-checking-module \citep{katoen2008principles,eker2004maude}.\\
Maude consists of a core which is named {\glqq Maude Core\grqq}. On top of its core every other part is written in Maude itself. Actually, Maude is distributed in version 2.6 form the website\footnote{\url{www.maude.cs.uiuc.edu/}, retrieval on 16/05/2014} \citep{clavel2011maude}.\\
\noindent A program in Maude is based on one or many modules where every part of the system looks like a clear to read abstract data type (ADT). A module contains a set of types, which are used with the {\glqq sort\grqq}-keyword. It is also possible to define more than one type with the plural form {\glqq sorts\grqq}. Each type describes a property for the module. For example types for a Petri net can be described with:

\begin{equation}
sort\text{ }Places\text{ }Transitions\text{ }Markings\text{ }.
\end{equation}

\noindent Depending on some sorts a set of operators needs to be defined. These operators describe all functors which are used to work with the defined types. For example a functor for writing a multiset of markings, can be expressed with a whitespace. This whitespace is surrounded with underscore, denoting a placeholder for the types defined after the double point. The return type right of the arrow, is of sort \textit{markings}.

\begin{equation}
op\text{ }\_\text{ }\_\text{ }:\text{ }Markings\text{ }Markings\text{ }\rightarrow \text{ }Markings\text{ }.
\end{equation}

\noindent If this operator has to be associative (in Maude with a short version: {\glqq assoc\grqq}) and commutative (short with: {\glqq comm\grqq}) properties, Maude defines this in the end of this line. Hence, we obtain a multiset of markings by this operator. The notation allows these properties in box brackets, so that it can be written as:

\begin{equation}
op\text{ }\_\text{ }\_\text{ }:\text{ }Markings\text{ }Markings\text{ }\rightarrow \text{ }Markings\text{ }[assoc\text{ }comm]\text{ }.
\end{equation}

\noindent Maude uses the equation logic to define the validity for an operator (axioms). This can be exemplified with the initial marking from a Petri net. This marking is a representation of the initial state of the Petri net. Based on this information we can define an operator that describes the initial state of a Petri net. After that, the validity with an equation can added. If we have a Petri net with only one marking with an {\glqq A\grqq} label we obtain these two lines:

\begin{equation}
op\text{ }initial\text{ }:\text{ }\rightarrow \text{ }Markings\text{ }.
\end{equation}
\begin{equation}
eq\text{ }initial\text{ }=\text{ }A\text{ }.
\end{equation}

\noindent Types are defined as {\glqq sort\grqq}, operators as functors and equations as the validity of operators. The rewrite rules can be used to replace one multiset with another multiset. So all terms are immutable as in many functional languages. $A$ replacement rule consists of two multisets, where the first set is replaced with the second one. These two termsets are separated with a double arrow, as shown in the following example, where a term $A$ is replaced with a new term $B$:

\begin{equation}\label{eq:TAB}
rl\text{ }[T]\text{ }:\text{ }A\text{ }\Rightarrow\text{ }B\text{ }.
\end{equation}

\noindent Based on this example an implementation of the token game of Petri nets can be realised. The two multisets can be seen as $pre$- and $post$-set of a transition. Hence, a rule can be used to describe a firing step with this two sets. This replacement rule can be modelled with the following graphical representation:

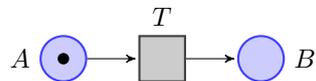
\begin{figure}[H]
\centering
\begin{tikzpicture}[node distance=1.3cm,>=stealth',bend angle=45,auto]
  \tikzstyle{place}=[circle,thick,draw=blue!75,fill=blue!20,minimum size=6mm]
  \tikzstyle{transition}=[rectangle,thick,draw=black!75,
  			  fill=black!20,minimum size=6mm]

  \tikzstyle{every label}=[black]

  \begin{scope}
    \node [place,tokens=1] (start) [label=left:$A$] {};

    \node [transition] (disable) [right of=start, label=above:$T$] {}
      edge [pre, left] (start);
    
    \node [place] (c) [right of=disable,label=right:$B$] {}
      edge [pre, left] (disable);
  \end{scope}
\end{tikzpicture}
\caption{Example Petri net $N$ for the \autoref{eq:TAB}}
\label{fig:petrinetzTAB}
\end{figure}

\subsection{Related Work}
The basic example for a Petri net to Maude conversion uses a shop system, where a user can buy an apple or candies. The mapping into Maude uses the term replacement system to model the firing steps of this net. Based on this Maude-structure it is possible to add a model-checking possibility, which can be used to verify a deadlock or safety properties \citep{clavel2011maude}.

\begin{figure}[H]
\centering
\includegraphics[keepaspectratio=true, width=.35\linewidth]{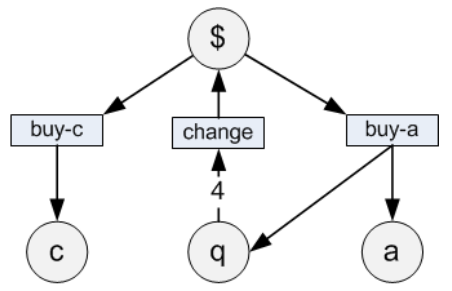}
\caption{Shop example with apple and candy}
\label{fig:appleCandy}
\vspace{-10pt}
\end{figure}

\noindent A more complex example of this modelling (see \autoref{fig:appleCandy}) is shown with high level nets in \citep{stehr2001rewriting}. It presents a conversion of the banker problem where two credits are handled by the net structure. The focus of the work lays on the soundness and correctness for this conversion approach. A formal definition of the model and the operators as well as the firing of a transition is given. Further, the paper presents an approach for a coloured Petri net (CPN). It extends the previous approach with more complex operators for the firing replacement rules. The conversion of transitions is realised via Maudes replacement rules and operators which contain the coloured tokens.\\
Automatic mapping for UML-models to a Maude-spe\-cification (see \citep{chama2013using}) is similar to this paper's idea of converting reconfigurable Petri nets. In \citep{chama2013using} the authors present three steps modeling, analysing and converting to Maude modules. The first step focuses on sub\-ject-specific modeling within UMLs class, state or components diagrams. After that step, the tool AtoM is used to convert the model into a Python-code representation. This code will be used to solve some constrains inside the UML-model components and some diagram specific parts. Lastly, the final step transfers all data into a Maude-specification, which can be used to verify some properties for example deadlocks.\\
In \citep{barbosa2011sysveritas} Petri nets are also converted into Maude-modules. As a base an Input-Output Place/Transition net (IOPT net) is used and saved in a PNML-file. These files are the origin for the conversion process. Further, PNML is used as a well-known markup-language for Petri nets. This process divides all components of a Petri net in special Maude-modules (net, semantic and initial markings) which can be used to verify in a same way as in \citep{chama2013using}.\\
A use case for the resulting Maude structure is presented in \citep{bjork2006executing}. The authors modelled the public transport of Oslo with a Petri net, which is converted into a Maude structure. The aim is to proof the net with Maude's LTL module and properties such as deadlock freedom or liveness. An evaluation shows that the simulation can handle a net with 2067 places and 3572 transitions as well up to 32 tokens (which models the trains) \citep{bjorkchallenges}. Test measurements for one train used 1614 rewrites and for 32 trains 35630 rewrites in 3.62 seconds.\\
\citep{boudiaf2009towards} presents a graphical editor for CPNs. It uses Maude in the background to verify properties such as liveness and deadlock freedom. Therefore, it converts a net into specified Maude modules (similar to \citep{stehr2001rewriting}) which are simulations with one step commands. After one step the tool is capable to present the results.

\newpage
\section{Data type}
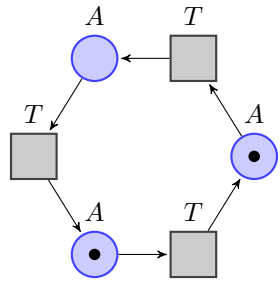
\begin{figure}[H]
\centering
\begin{tikzpicture}[node distance=1.3cm,>=stealth',bend angle=45,auto]
  \tikzstyle{place}=[circle,thick,draw=blue!75,fill=blue!20,minimum size=6mm]
  \tikzstyle{transition}=[rectangle,thick,draw=black!75,
  			  fill=black!20,minimum size=6mm]

  \tikzstyle{every label}=[black]

  \begin{scope}
    \node [place] (A1) [label=above:$A$] {};

    \node [transition] (T1) [right of=A1, label=above:$T$] {}
      edge [post, above] (A1);

    \node [place,tokens=1] (A2) [xshift=8mm, below of=T1, label=above:$A$] {}
      edge [post, above] (T1);
      
    \node [transition] (T2) [xshift=-8mm, below of=A2, label=above:$T$] {}
      edge [post, above] (A2);
      
    \node [place,tokens=1] (A3) [left of=T2, label=above:$A$] {}
      edge [post, above] (T2);
      
    \node [transition] (T3) [xshift=-8mm, above of=A3, label=above:$T$] {}
      edge [pre, above] (A1)
      edge [post, above] (A3);
  \end{scope}
\end{tikzpicture}
\caption{Example Petri net $N_{1}$}
\label{fig:petrinetz}
\end{figure}

A reconfigurable Petri net $N_{1}$ consists of a tuple which is separated in a Petri net $N$ and a set of rules $\cal R$. It can be written with $N_{1} = (N, {\cal R})$. Furthermore, a Petri net can be formally described as a tuple $N = (P, T, pre, post, M, cap)$. Where $P$ is a set of places, $T$ is a set of transitions, $pre$ and $post$ are functions which maps $T \rightarrow P^{\oplus}$ and finally $M$ is the initial marking. Additionally, a function $cap : P \rightarrow N^{\omega}$ can be used to model a capacity of a place $P$ with a value $N^{\omega}$ \citep{padberg2012abstract}.\\
An example of a Petri net is shown in \autoref{fig:petrinetz}. Each blue circle is a place and each rectangle is a transition. The arrows between these elements describe the arcs, which can connect a place with a transition and vice versa. The two black points are tokens which can be consumed by a transition.

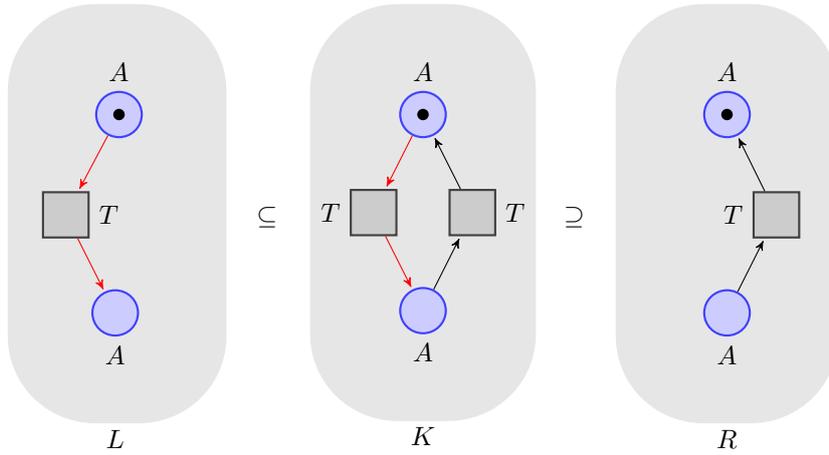
\begin{figure}[H]
\centering
\begin{tikzpicture}[node distance=1.3cm,>=stealth',bend angle=45,auto]
  \tikzstyle{place}=[circle,thick,draw=blue!75,fill=blue!20,minimum size=6mm]
  \tikzstyle{transition}=[rectangle,thick,draw=black!75,
     fill=black!20,minimum size=6mm]

  \tikzstyle{every label}=[black]

  \begin{scope}
    \node [place,tokens=1] (A1_l) 
      [label=above:$A$] {};
	    
    \node [transition] (T_l) 
      [yshift=-1.33cm, xshift=-20mm, right of=A1_l, label=right:$T$] {}
      edge [pre, right, color=red] (A1_l);	  
	  
    \node [xshift=1.35cm, yshift=0.35cm, right of=T_l, label=below:$\subseteq$] {}; 
	    
	\node [place] (A2_l) 
	  [xshift=6.5mm, below of=T_l, label=below:$A$] {}
      edge [pre, right, color=red] (T_l);
	
	\node [below of=A2_l, label=below:$L$] {};
  \end{scope}
  
  \begin{scope}[xshift=4cm]
    \node [place,tokens=1] (A1_k) 
      [label=above:$A$] {};

    \node [transition] (T1_k) 
      [xshift=-6.5mm, below of=A1_k, label=left:$T$] {}
      edge [pre, left, color=red] (A1_k);
      
    \node [transition] (T2_k) 
      [right of=T1_k, label=right:$T$] {}
      edge [post, right] (A1_k);

    \node [place] (A2_k) 
      [xshift=6.5mm, below of=T1_k, label=below:$A$] {}
      edge [post, left] (T2_k)
      edge [pre, left, color=red] (T1_k);

	\node [below of=A2_k, label=below:$K$] {};
  \end{scope}
  
  \begin{scope}[xshift=8cm]
    \node [place,tokens=1] (A1_r) 
      [label=above:$A$] {};
	    
    \node [transition] (T_r) 
      [yshift=-1.33cm, xshift=-6.5mm, right of=A1_r, label=left:$T$] {}
      edge [post, right] (A1_r);	  
	  
    \node [yshift=0.35cm, xshift=-1.36cm, left of=T_r, label=below:$\supseteq$] {};  
	    
	\node [place] (A2_r) 
	  [xshift=-6.5mm, below of=T_r, label=below:$A$] {}
      edge [post, right] (T_r);
	
	\node [below of=A2_r, label=below:$R$] {};
  \end{scope}
    
  \begin{pgfonlayer}{background}
    \filldraw [line width=23mm, join=round, black!10]      
      (A1_l.north  -| A2_l.east)  rectangle 
      (A2_l.south  -| A1_l.west);      
    \filldraw [line width=23mm, join=round, black!10]      
      (A1_k.north  -| T1_k.east)  rectangle 
      (A2_k.south  -| T2_k.west);
    \filldraw [line width=23mm, join=round, black!10]      
      (A1_r.north  -| A2_r.east)  rectangle 
      (A2_r.south  -| A1_r.west);      
  \end{pgfonlayer}
\end{tikzpicture}
\caption{Example rule $r_{1}$, which switch the arc direction}
\label{fig:exampleRule}
\end{figure}

\noindent A rule is consists of three Petri nets $L, K$ and $R$. $L$ is the left-hand side (LHS) which should be found. The right-hand side $R$ would be inserted in the net, if this rule is used. $K$ is the interface between $L$ and $R$.\\
The example in \autoref{fig:exampleRule} shows a rule which changes the direction of an arc for a transition $T$. The change is realised by two steps. At first, the match of the left-hand side ensures that the rules can be applied. And finally, the right-hand side contains the information to be used.\\
This example contains a transition which connects the places in reverse order (arc colour black). However, the mapping net contains both transitions. The arc inversion is realised by a deleting one transition and adding a new transition with reversed arcs.

\noindent The aim of this work is to create Maude modules, which provides the possibility to create a formal writing of a reconfigurable Petri net. The example net in \autoref{fig:petrinetz} can be formally indicated as:

\begin{figure}[H]
\begin{eqnarray*}
\begin{aligned}[c]
P &=& \{A_{2}, A_{3}, A_{4}\}\\
T &=& \{T_{5}, T_{6}, T_{7}\}\\
pre(T_{5}) &=& A_{4}\\
pre(T_{6}) &=& A_{2}\\
pre(T_{7}) &=& A_{3}\\
\end{aligned}
\qquad \qquad
\begin{aligned}[c]
post(T_{5}) &=& A_{3}\\
post(T_{6}) &=& A_{4}\\
post(T_{7}) &=& A_{2}\\
m &=& A_{2} + A_{4}\\
cap &=& \omega A_{2} + \omega A_{3} + \omega A_{4}
\end{aligned}
\end{eqnarray*}
\caption{Formal description of a Petri net (for a graphical presentation see also \autoref{fig:petrinetz})}
\label{eq:formalDef}
\end{figure}

\noindent Hence, the module needs a definition for places, transitions and markings as well as the definition of pre- and post-sets. It comprised the type hierarchy and the syntax as shown in \autoref{lst:pnDef}.\\
First, all sorts are defined. A \textit{sort} can be understand as a type which contains the semantic. An example for a semantic is a Petri net place or transition. Furthermore, a type hierarchy can be formulated with the keyword \textit{subsort}. This is a membership equation logic feature of Maude. It enables Maude to use a mapping between different types. This facilitated to describe the type \textit{Places} as subset of the type \textit{Markings}.\\
Based on the types, all operators can be defined. A definition begins with the keyword \textit{op} followed by the operator syntax. This contains an operator-name and handover parameter. For all parameters a definition of the types are shown after the colon. If this multi-set of parameters is empty (an arrow stands after the colon) this operator is a constant. The return type stands after the arrow and finally a dot ends the statement. Additionally, special properties as associativity or commutativity can be added inside of box brackets. Further, the definition can be extended with equations.

\begin{minipage}[H]{0.95\linewidth}
\centering
\begin{lstlisting}[breaklines=true, language=maude, caption={Maude module for a Petri net}, label={lst:pnDef}, mathescape=true]
  sorts Net Places Transitions Pre Post MappingTuple Markings .

  subsort Places < Markings .

  op emptyPlace : -> Places .
  op emptyTransition : -> Transitions .
  op emptyMappingTuple : -> MappingTuple .
  op emptyMarking : -> Markings .

  op _,_ : Places Places -> Places [ctor assoc comm id: emptyPlace] .
  op _+_ : Places Places -> Places [ctor assoc comm id: emptyPlace] .
  op _:_ : Transitions Transitions -> 
            Transitions [ctor assoc comm id: emptyTransition] .
  op _,_ : MappingTuple MappingTuple -> 
            MappingTuple [ctor assoc comm id: emptyMappingTuple] .
  op _;_ : Markings Markings -> Markings [ctor assoc comm id: emptyMarking] .

  *** READING: Pname | ID | Cap
  op p(_|_|_) : String Int Int -> Places .
  op t(_|_) : String Int -> Transitions .
  op (_-->_) : Transitions Places -> MappingTuple .

  op places{_} : Places -> Places .
  op transitions{_} : Transitions -> Transitions .
  op pre{_} : MappingTuple -> Pre .
  op post{_} : MappingTuple -> Post .
  op marking{_} : Markings -> Markings .

  *** Petrinet-tuple
  op net : Places Transitions Pre Post Markings -> Net .
\end{lstlisting}
\end{minipage}

\noindent Based on this definition, it is possible to write the net in Maude. The operator \textit{net} is a wrapper-operator for a net and contains \textit{Places, Transitions, Pre, Post} and \textit{Markings}. The set of places is realised with the operator \textit{places}. This operator contains a multi-set for places which are separated with a comma. Further, the operator \textit{transitions} is a set for transitions. It separates the elements with a colon. In addition, the \textit{pre} and \textit{post} operators describe the pre and post conditions of a transition. Both operators contain a multi-set of \textit{MappingTuple}, which are a mapping between a transition and a multi-set of places. Finally, the \textit{marking}-operator contains a multi-set of \textit{Places}. The content is separated with an additional symbol as the linear sum in the mathematics definition in \autoref{eq:formalDef}.\\
The example in \autoref{eq:formalDef} can be written in Maude as in \autoref{lst:pnExampleMaudeCode}. It separates all operators and the included multi-sets with commas. Each multi-set is wrapped with curved brackets. A place is modelled as tuple with \textit{p($<$label$>$ $|$ $<$identifier$>$ $|$ $<$capacity$>$)}. Labels are defined as a string, identifier and capacity as numbers. Transitions are based on the tuple \textit{t($<$label$>$ $|$ $<$identifier$>$)}. Each type has the same type as for the place.

\begin{minipage}[H]{0.95\linewidth}
\centering
\begin{lstlisting}[breaklines=true, language=maude, caption={Maude module for a Petri net}, label={lst:pnExampleMaudeCode}, mathescape=true]
net(places{ p("A" | 3 | 2147483647) , p("A" | 4 | 2147483647) ,
             p("A" | 2 | 2147483647) } ,
    transitions{ t("T" | 7) : t("T" | 5) : t("T" | 6) } ,
    pre{ (t("T" | 7) --> p("A" | 3 | 2147483647)) ,
          (t("T" | 5) --> p("A" | 4 | 2147483647)) , 
          (t("T" | 6) --> p("A" | 2 | 2147483647)) } ,
    post{ (t("T" | 7) --> p("A" | 2 | 2147483647)) ,
           (t("T" | 5) --> p("A" | 3 | 2147483647)) , 
           (t("T" | 6) --> p("A" | 4 | 2147483647)) } ,
    marking{ p("A" | 3 | 2147483647) ; p("A" | 4 | 2147483647) } )
\end{lstlisting}
\end{minipage}

\subsection{Activation and Firing}
A transition $t$ is activated, written by $m[t\rangle$, when the following two conditions are satisfied. The first condition consists of the pre-set of this transition. The net marking has to contain at least as many tokens, as described it in the pre-set (see \autoref{mat:activation}). Furthermore, all post places have to satisfy the capacity condition. Adding more tokens than a place can store is not possible (see \autoref{mat:cap}).\\
If both conditions are satisfied, the transition $t$ can fire. One firing step is written with $m [t\rangle m'$, where $m$ is the current marking and $m'$ is the following marking. The calculation of $m'$ is described in \autoref{mat:followingMarking}. First, the pre-set is deducted from the current marking. Now the post-set of $t$ can be added to the result.

\begin{equation}\label{mat:activation}
pre^{\oplus}(t) \leq m
\end{equation}
\begin{equation}\label{mat:cap}
m + post^{\oplus}(t) \leq cap
\end{equation}
\begin{equation}\label{mat:followingMarking}
m' = (m \ominus pre^{\oplus} (t)) \oplus post^{\oplus} (t).
\end{equation}

\noindent A conversion of the first condition (see \autoref{mat:activation}) is shown in \autoref{lst:activateAndFire}. The rewrite rule contains two parts for the condition. First, \textit{T --$>$ PreValue} models \textit{pre$^{\oplus}$(t)}. Second, $\leq m$ is converted into \textit{marking\{PreValue ; M\}}. Hence, the formal definition is implemented. Either the pre-set of a transition is a part of the marking multi-set, or the rule is not enabled for firing. This implementation uses the matching algorithm from Maude to find possible cases of applications. It is able to determine when the termset contains this condition. In summary, a rule uses one transition from the \textit{net}-tuple and tests the existing in the \textit{pre}-set in the current marking.\\
Furthermore, the \autoref{mat:cap} is expressed by the condition of the Maude rule. Hence, the sum of the current marking plus the post-set for the transition is less or equal than the capacity of each place. The addition of the current marking and the post-set is written after the \textit{if} in the last line of \autoref{lst:activateAndFire}. The addition result is used with the $<=?$ which requires a multi-set of places on the right side. Details can be found in \autoref{lst:activateAndFireCondition2}.\\
Further, the rule result contains a function which calculates the resulting set of markings.

\[ \text{\textit{calc}}(((\text{\textit{PreValue }};\text{\textit{ M}})\text{\textit{ minus PreValue}})\text{\textit{ plus PostValue}}) \]

\noindent And may be read as the formal definition in \autoref{mat:followingMarking}, where \textit{PreValue ; M} describes \textit{m} and \textit{PreValue} is the pre-value of the transition \textit{v}. The place holder \textit{PostValue} represents the post-domain of the transition \textit{t}.

\begin{minipage}[H]{0.95\linewidth}
\centering
\begin{lstlisting}[breaklines=true, language=maude, caption={Activation and firing of a transition}, label={lst:activateAndFire}, mathescape=true]
crl [fire] : 
     net(P,
         transitions{T : TRest}, 
         pre{(T --> PreValue), MTupleRest1}, 
         post{(T --> PostValue), MTupleRest2}, 
         marking{PreValue ; M}) 
     Rules
     I
     =>
     net(P, 
         transitions{T : TRest}, 
         pre{(T --> PreValue), MTupleRest1}, 
         post{(T --> PostValue), MTupleRest2}, 
         calc(((PreValue ; M) minus PreValue) plus PostValue))
     Rules
     I
     if calc((PreValue ; M) plus PostValue) <=? PostValue .
\end{lstlisting}
\end{minipage}

\noindent The operator $<=?$ (in words \textit{smallerAsCap}) maps the capacity condition in this Maude module. The aim is to return \textit{true} if the marking is less or equal than the capacity of each places in the post-set of a transition.\\
The source code consists of the helper method \textit{\_ leeqth \_ with \_}. This operator tests the capacity of a place multi-set and a single place. The third parameter is used to count the occurrence of a token in the multi-set. In case that the counter is bigger than the capacity it returns false. All other cases results with true.

\begin{minipage}[H]{0.95\linewidth}
\centering
\begin{lstlisting}[breaklines=true, language=maude, caption={Capacity proof of each place in the post-set}, label={lst:activateAndFireCondition2}, mathescape=true]
  op _ <=? _ : Markings Places -> Bool .
  op _ leeqth _ with _ : Places Places Int -> Bool .

  *** Impl - smallerAsCap #############
  eq marking{ PSet } <=? emptyPlace = true .
  ceq marking{M} <=? (P , emptyPlace)
      = true
      if M leeqth P with 0 .
  ceq marking{M} <=? (P , PRest)
      = true
      if M leeqth P with 0
         /\ PRest =/= emptyPlace
         /\ marking{M} <=? PRest .
  eq M <=? P = false [owise] .
  
  *** Impl - lowerEqualThan ##########
  *** place multiset is empty
  ceq emptyMarking leeqth p(Str | I | Cap1) with Counter
      = true if Counter <= Cap1 .
  *** Cap-counter is too big
  eq (p(Str | I | Cap2) ; MRest) leeqth p(Str | I | Cap2) with (Cap2 + 1) 
     = false .
  *** found same place
  ceq (p(Str | I | Cap2) ; MRest) leeqth p(Str | I | Cap2) with Counter 
      = true 
      if (MRest leeqth p(Str | I | Cap2) with (Counter + 1)) .
  *** del another place
  ceq (p(Str | I | Cap1) ; MRest) leeqth p(Str2 | I2 | Cap2) with Counter 
      = true 
      if (MRest leeqth p(Str2 | I2 | (Cap2)) with Counter) .
  *** otherwise
  eq M leeqth P with I = false [owise] .
\end{lstlisting}
\end{minipage}

\subsection{LTL Properties}
The aim of this work is to verify properties such as deadlocks, liveness or reachability for a reconfigurable Petri net. To realise this Maude's LTLR implementation is used. It is based on an implementation of the linear temporal logic (LTL). The implementation itself uses a Kripke structure, which is realised on the basis of the equation and rewriting logic, basically a finite transition system \citep{eker2004maude}.\\
The following examples in \autoref{lst:enabledOfN1} ff. are using the operators defined in \autoref{lst:LTLProp}. It contains an operator for the reachability of a marking. The \textit{enabled}-operator includes the activation of a transition as well as the ability to apply a rule. Finally, the last three lines in \autoref{lst:LTLProp} include a standard equation, which is used when no other equation can be used.\\
The implementation starts with a sub sorting of the \textit{Configuration}-type. This is necessary because the Kripke structure is based on these informations. It means that all \textit{Configuration}-objects are relevant for the construction of the states of the Kripke structure. In terms of this work a \textit{Configuration}-object contains a snapshot of a reconfigurable Petri net. At the beginning it includes the initial marking and the primal state of the net without any transformation with rules. All other following conditions include the further interactions of the net (in this case the \textit{Configuration} with firing steps and rule treatments).

\begin{lstlisting}[breaklines=true, language=maude, caption={LTL Properties: deadlocks, liveness and reachability}, label={lst:LTLProp}, mathescape=true]
subsort Configuration < State .

op reachable : Markings -> Prop .

eq net(P , T , Pre , Post ,
        marking{ M ; MRest } ) 
        Rules MaxID StepSize aidP aidT
    |= reachable(M) = true .

op t-enabled : -> Prop .

eq net(P , T ,
        pre{ (T1 --> PreValue) , MappingTuple } ,
        Post ,
        marking{ PreValue ; MRest } ) 
        Rules MaxID StepSize aidP aidT 
    |= t-enabled = true .
  eq C |= t-enabled = false [owise] . 

op enabled : -> Prop .

eq net(P , T ,
        pre{ (T1 --> PreValue) , MappingTuple } ,
        Post ,
        marking{ PreValue ; MRest } ) 
        Rules MaxID StepSize aidP aidT
    |= enabled = true .

eq net(places{ p("A" | Irule2017 | 2147483647) ,
                p("A" | Irule2020 | 2147483647) , P } ,
        transitions{ t("T" | Irule2024) : T } ,
        pre{ (t("T" | Irule2024) --> p("A" | Irule2017 | 2147483647)) , 
               MTupleRest1 } ,
        post{ (t("T" | Irule2024) --> p("A" | Irule2020 | 2147483647)) ,
               MTupleRest2 } ,
        marking{ p("A" | Irule2017 | 2147483647) ; M } )
        Rules MaxID StepSize aidP aidT
    |= enabled = true .

var C : Configuration .
var Prop : Prop .
eq C |= Prop = false [owise] .
\end{lstlisting}

\subsection{Matching of Rules}
A reconfigurable Petri net consists of a net and a set of rules $\cal R$. Each rule contains three sub-nets, which contain a net $L$ for matching, a net $R$ for the replacement and a net $K$ that maps between the two nets.\\
The first project \citep{project1} is based on \textit{ReConNet}. This tool provides the capability to find non-deterministic matches for a net and a set of rules \citep{Blumreiter13}. The aim of the first project is an extension that enables \textit{ReConNet} to verify a net with a given set of rules. The verification process is realised by a conversion to a Maude specification. In order to realise this process, an interface is designed for using \textit{ReConNet} to find a match. 
Due to this constellation, only the initial state of a net and all rules can be verified with the LTL-process.\\
The new aim is to ensure that the Maude specification finds the matching itself. This implies a possibility to define a rule in this specification as well as the dangling-condition (see section \ref{sec:dangling}). Further, the meta-data configuration (such as current highest identifier) should adapt a net and a set of rules.\\
The definition of a rule and the meta configuration can be found in \autoref{lst:RuleAndConfiguration}. First, the sorts \textit{Rule, LeftHandSide} and \textit{RightHandSide} are defined. This models the two sides of a rule. The mapping net $K$ is not included, because it is not relevant for matching of a rule. Further, the \textit{Configuration} consists of a net, a multi-set of rules and a global ID-count. The net contains all information as places, transitions, markings and pre- as well as post-sets. Each rule multi-set entry contains a left and right side. At last the ID-count is used for each insertion step, where a new transition or a place will be added.

\begin{minipage}[H]{0.95\linewidth}
\centering
\begin{lstlisting}[breaklines=true, language=maude, caption={Definition of a rule and the configuration (net, multi-set of rules and global ID-count)}, label={lst:RuleAndConfiguration}, mathescape=true]
*** Rule R = (l_net, r_net)
sort Rule .
sorts LeftHandSide RightHandSide .
op emptyRule : -> Rule .
op _|_ : Rule Rule -> Rule [ctor assoc comm id: emptyRule] .
op l : Net -> LeftHandSide .
op r : Net -> RightHandSide .
op rule : LeftHandSide RightHandSide -> Rule .

*** Configuration
sort Configuration .
op ___ : Net Rule Int -> Configuration .
\end{lstlisting}
\end{minipage}

\noindent The rule $r_{1}$ is shown in two Listings \ref{lst:exampleRuleLHS} and \ref{lst:exampleRuleRHS}. It is based on the definition in \autoref{lst:RuleAndConfiguration}. Where the left-hand side provides all information which are necessary for finding a match. The right-hand side contains all information for the result of the transformation. For example it contains the new elements (here a transition) and the related ID.\\
The whole rule is written as a condition replacement. If the conditions satisfied with the current net, the rule can be applied. Due to the replacement, the left-hand side of the rule is replaced with the right-hand side. Furthermore, a rule is working with a configuration object. It uses a net, a multi-set of rules as well as an ID-count.\\
The example in \autoref{lst:exampleRuleLHS} shows the left side of the rule in \autoref{fig:exampleRule}. The net which should be found, consists of two places with the label $A$. Further, it contains a transition $T$ which connects the two places. A description of the arcs can be found in the pre- and post-set. At last it contains a marking on the place $A$. All these elements are part of the net in the rule. It only differs in the ID, which are not given as concrete numbers. Each \textit{Irule$<$number$>$} is a variable that is used for finding a structural match. If the net has the same structure, but other ID's for the places and transitions, this rule is also activated. This is possible because Maude maps the ID's internal with each variable. Moreover, each set contains a variable for possible residual elements. For example, in this net the set $places$ contains more than two places. The remaining place's are mapped into the variable $PRest$, so this rule is still activated. If a multiset has no more elements the identities (id) is used (for a definition see \autoref{lst:pnDef}). Each id itself is the empty-set constant operator (such as emptyPlace, emptyTransition, emptyMappingTuple or emptyMarking).

\begin{minipage}[H]{0.95\linewidth}
\centering
\begin{lstlisting}[breaklines=true, language=maude, caption={Example of rule $r_{1}$ written with Maude (left-hand side)}, label={lst:exampleRuleLHS}, mathescape=true]
crl [R1-PNML] :
     net(places{p("A" | Irule1017 | 2147483647) , 
                  p("A" | Irule1020 | 2147483647) , PRest} ,
          transitions{ t("T" | Irule1024) : TRest} ,
          pre{(t("T" | Irule1024) --> p("A" | Irule1017 | 2147483647)) ,
                 MTupleRest1} ,
          post{(t("T" | Irule1024) --> p("A" | Irule1020 | 2147483647)) ,
                  MTupleRest2} ,
          marking{p("A" | Irule1017 | 2147483647) ; MRest} )
     rule(l(net(places{p("A" | Irule2017 | 2147483647) , 
                        p("A" | Irule2020 | 2147483647)} ,
                 transitions{t("T" | Irule2024)} ,
                 pre{(t("T" | Irule2024) --> 
                         p("A" | Irule2017 | 2147483647))} ,
                 post{(t("T" | Irule2024) --> 
                          p("A" | Irule2020 | 2147483647))} ,
                 marking{p("A" | Irule2017 | 2147483647)} ) ) ,
          r(net(places{p("A" | Irule3017 | 2147483647) , 
                        p("A" | Irule3020 | 2147483647)} ,
                 transitions{t("T" | Irule3026)} ,
                 pre{(t("T" | Irule3026) --> 
                         p("A" | Irule3020 | 2147483647))} ,
                 post{(t("T" | Irule3026) --> 
                          p("A" | Irule3017 | 2147483647))} ,
                 marking{p("A" | Irule3017 | 2147483647)} ) ) )
     | RRest MaxID StepSize 
     aidPlace{AidPRest} aidTransition{AidTRest}
\end{lstlisting}
\end{minipage}

\noindent In the following example in \autoref{lst:exampleRuleRHS} the right-hand side differs from the left-hand side. It contains the net structure of the right-hand side. The example adds the new transition $T$. It calculates the new identifier which is detailed described in \autoref{sec:identifierKrams}.

\begin{minipage}[H]{0.95\linewidth}
\centering
\begin{lstlisting}[breaklines=true, language=maude, caption={Example of the rule $r_{1}$, written with Maude (right-hand side)}, label={lst:exampleRuleRHS}, mathescape=true]
     =>
     net(places{p("A" | Irule1017 | 2147483647) , 
                 p("A" | Irule1020 | 2147483647) , PRest} ,
          transitions{t("T" | AidT1 ) : TRest} ,
          pre{(t("T" | AidT1 ) --> p("A" | Irule1020 | 2147483647)) , 
                MTupleRest1} ,
          post{(t("T" | AidT1 ) --> p("A" | Irule1017 | 2147483647)) ,
                 MTupleRest2} ,
          marking{p("A" | Irule1017 | 2147483647) ; MRest} )
     rule(l(net(places{p("A" | Irule2017 | 2147483647) , 
                        p("A" | Irule2020 | 2147483647)} ,
                 transitions{t("T" | Irule2024)} ,
                 pre{(t("T" | Irule2024) --> 
                          p("A" | Irule2017 | 2147483647))} ,
                 post{(t("T" | Irule2024) --> 
                          p("A" | Irule2020 | 2147483647))} ,
                 marking{p("A" | Irule2017 | 2147483647)} ) ) ,
           r(net(places{p("A" | Irule3017 | 2147483647) , 
                         p("A" | Irule3020 | 2147483647)} ,
                  transitions{t("T" | Irule3026)} ,
                  pre{(t("T" | Irule3026) --> 
                        p("A" | Irule3020 | 2147483647))} ,
                  post{(t("T" | Irule3026) --> 
                           p("A" | Irule3017 | 2147483647))} ,
                  marking{p("A" | Irule3017 | 2147483647)} ) ) )
     | RRest NewMaxID StepSize 
     aidPlace{AidPRest} aidTransition{AidTRest2}
     if AidTRest1 := addOldID(AidTRest | Irule1024) /\
        AidT1 := getAid(AidTRest1 | MaxID | StepSize) /\
        AidTRest2 := removeFirstElement(AidTRest1 | MaxID | StepSize) /\
        NewMaxID := correctMaxID(MaxID | StepSize | 2) .
\end{lstlisting}
\end{minipage}

\subsection{Dangling-Condition}\label{sec:dangling}
\begin{figure}[H]
\begin{minipage}{.35\textwidth}
\centering
\begin{tikzpicture}[node distance=1.3cm,>=stealth',bend angle=45,auto]
  \tikzstyle{place}=[circle,thick,draw=blue!75,fill=blue!20,minimum size=6mm]
  \tikzstyle{transition}=[rectangle,thick,draw=black!75,
     fill=black!20,minimum size=6mm]

  \tikzstyle{every label}=[black]

  \begin{scope}
    \node [place, tokens=1] (A1_l)
      [label=above:$A$] {};

    \node [transition] (T_l)
      [right of=A1_l, label=above:$T$] {}
      edge [pre, right] (A1_l);

	\node [place] (A2_l) 
	  [right of=T_l, label=above:$A$, color=red] {}
      edge [pre, right] (T_l);
      
    \node [place, tokens=1] (A1_l_2)
      [below of=A1_l, label=above:$A$] {};

    \node [transition] (T_l_2)
      [right of=A1_l_2, label=above:$T$] {}
      edge [pre, right] (A1_l_2);

	\node [place] (A2_l_2) 
	  [right of=T_l_2, label=above:$A$] {}
      edge [pre, right] (T_l_2);

    \node [transition] (T_l2_2)
      [right of=A2_l_2, label=above:$T$] {}
      edge [pre, right] (A2_l_2);
  \end{scope}
\end{tikzpicture}
\caption{Example net $N_{2}$}
\label{fig:exampleNetDangling}
\end{minipage}
\quad
\begin{minipage}{.65\textwidth}
\begin{figure}[H]
\centering
\begin{tikzpicture}[node distance=1.3cm,>=stealth',bend angle=45,auto, scale=0.8, transform shape]
  \tikzstyle{place}=[circle,thick,draw=blue!75,fill=blue!20,minimum size=6mm]
  \tikzstyle{transition}=[rectangle,thick,draw=black!75,
     fill=black!20,minimum size=6mm]

  \tikzstyle{every label}=[black]

  \begin{scope}
    \node [place, tokens=1] (A1_l)
      [label=above:$A$] {};

    \node [transition] (T_l)
      [below of=A1_l, label=left:$T$] {}
      edge [pre, right] (A1_l);

    \node [xshift=0.7cm, yshift=0.35cm, right of=T_l, label=below:$\subseteq$] {}; 

	\node [place] (A2_l) 
	  [below of=T_l, label=below:$A$] {}
      edge [pre, right] (T_l);

	\node [below of=A2_l, label=below:$L$] {};
  \end{scope}
  
  \begin{scope}[xshift=4cm]
    \node [place, tokens=1] (A1_k) 
      [label=above:$A$] {};

    \node [transition] (T1_k) 
      [below of=A1_k, label=left:$T$] {}
      edge [pre, left] (A1_k);

    \node [place] (A2_k) 
      [below of=T1_k, label=below:$A$] {}
      edge [pre, left] (T1_k);

	\node [below of=A2_k, label=below:$K$] {};
  \end{scope}

  \begin{scope}[xshift=8cm]
    \node [place, tokens=1] (A1_r)
      [label=above:$A$] {};

    \node [yshift=-9.5mm, xshift=-0.7cm, left of=A1_r, label=below:$\supseteq$] {};

	\node [transition] (A2_r) 
	  [below of=A1_r, label=left:$T$] {}
	  edge [pre, left] (A1_r);

	\node [yshift=-13mm, below of=A2_r, label=below:$R$] {};
  \end{scope}

  \begin{pgfonlayer}{background}
    \filldraw [line width=17mm, join=round, black!10]
      (A1_l.north  -| A2_l.east)  rectangle
      (A2_l.south  -| A1_l.west);
    \filldraw [line width=17mm, join=round, black!10]
      (A1_k.north  -| T1_k.east)  rectangle
      (A2_k.south  -| A2_k.west);
    \filldraw [line width=17mm, join=round, black!10]
      (A1_r.north  -| A2_r.east)  rectangle
      (A2_k.south  -| A1_r.west);
  \end{pgfonlayer}
\end{tikzpicture}
\end{figure}
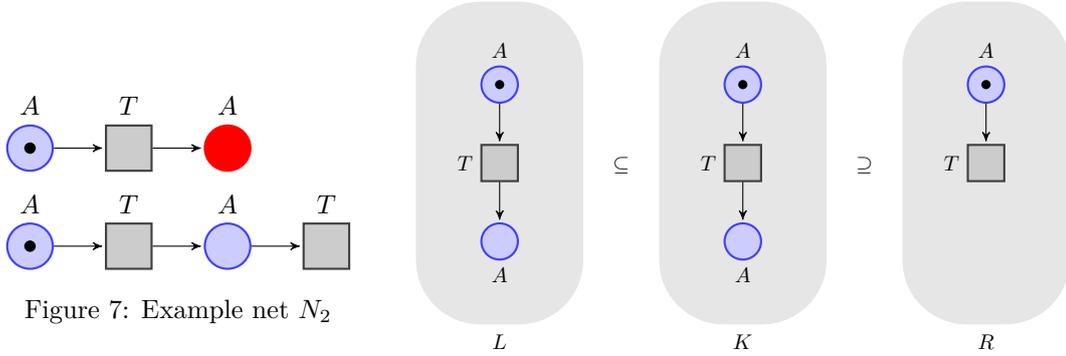
\caption{Example rule $r_{3}$ which deletes the place \textit{A}}
\label{fig:exampleRuleDangling}
\end{minipage}
\end{figure}

A special part of a rule matching is the gluing condition. This condition is separated into the identification and dangling condition. The identification condition requires that no place or transition is specified to be simultaneously added and deleted. Further, the dangling condition defines that a place can only be deleted if there are only arcs to transitions, that are deleted as well. Transitions are not relevant for dangling condition.\\
The example net $N_{2}$ in \autoref{fig:exampleNetDangling} shows a short example, where a place $A$ (the red place) should be deleted with rule $r_{3}$ in \autoref{fig:exampleRuleDangling}. The rule has only one match because the bottom part of the net differs in an addition transition. This transition injured the dangling condition and the rule cannot be used at this point.

\noindent The associated Maude code can be found in \autoref{lst:danglingCondition}. It is seperated in two parts. First, the \textit{equalMarking}-operator tests the current net marking. This condition ensures that the net marking and the rule are the same for all deleted places.\\
The marking of the net is given with:

\[ \text{marking}\{\text{ p(\glqq}\text{A}\text{\grqq }|\text{ Irule1019 }|\text{ 2147483647})\text{ ; MRest }\} \]

\noindent It contains the token $A_{Irule1019}$ from the rule and also the variable \textit{MRest}. The second marking, from the net (another $A$ token), is in this multi-set of markings. Further, the second multi-set contain all places that should be deleted. In this example the other place is called \textit{A}. Hence, the following line contains all relevant information:

\[ \text{marking}\{\text{ p(\glqq A\grqq }|\text{ Irule2019 }|\text{ 2147483647) }\} \]

\noindent The result is true, if the second set of markings is equal to the marking in the first set.\\
Furthermore, for each deleted place is a \textit{emptyNeighbourForPlace}-operator defined. This operator tests the dangling condition with the remaining pre and post multi-sets. In \autoref{lst:danglingCondition} this two sets are defined as \textit{MTupleRest1} and \textit{MTupleRest2}. All operator definitions can be found in \autoref{lst:danglingHelper}.

\begin{minipage}[H]{0.95\linewidth}
\centering
\begin{lstlisting}[breaklines=true, language=maude, caption={Dangling-Condition if a place should be deleted}, label={lst:danglingCondition}, mathescape=true]
equalMarking( ( marking{p("S" | Irule107 | 2147483647) ; MRest} )
                  =?=
                  marking{emptyMarking} ) /\
emptyNeighbourForPlace(p("S" | Irule106 | 2147483647) ,
                          pre{ MTupleRest1 } ,
                          post{ MTupleRest2 } )
\end{lstlisting}
\end{minipage}

\noindent All needed operators for the dangling condition are defined in \autoref{lst:danglingHelper}. The \textit{equalMarking}-operator proves whether the second multi-set is a subset of the first multi-set. The first parameter is a multi-set of the current net marking. And the second parameter contains the marking of the rule. The result is only true, if the first multi-set contains the same marking for each place as the second multi-set. Therefore, the equations are separated into five case differentiations. The first equation includes the situation where two identical marks are compared. If this situation occurs the result is true. The second case contains a recursion. It consumes two markings of each multi-set and returns true, if the recursion call also returns true. The following equation differs only in case distinction. It returns true, including a recursion, when the markings are not in the multi-sets. Based on this equation, the following lines include the situation where the second set only contains one element. If the first remaining multi-set does not contain more than one of these markings, it returns true. Finally, if it is not possible to use another case, this is called the {\glqq otherwise case\grqq} (Maude's keyword with brackets: \textit{[owise]}). For all other conditions it returns false and ends the recursion.\\
Further, the \textit{emptyNeighbourForPlace}-operator can be used for a test, that examines the neighbours of a place. It returns true if the place does not have any arcs outside the rule. This means exactly that all arcs at the place are included in the rule.\\
The implementation uses three equations. They are separated into the first two situations where a pre or a post exists. And finally the other case is true, where no arc exists.

\begin{lstlisting}[breaklines=true, language=maude, caption={Dangling-Condition helper methods}, label={lst:danglingHelper}, mathescape=true]
op contains(_ | _) : Places Places -> Bool .
eq contains(p(Str | I | Cap) | (p(Str | I | Cap), PRest)) = true .
eq contains(P | PNet) = false [owise] .

*** READING: NET-MARKING, RULE-MARKING
op equalMarking(_ =?= _) : Places Places -> Bool .
eq equalMarking(p(Str | I1 | Cap) =?= p(Str | I2 | Cap)) = true .
ceq equalMarking(
    (p(Str | I1 | Cap) , MNet) =?= (p(Str | I2 | Cap) , MRest))
    = true
    if contains((p(Str | I1 | Cap)) | MNet) /\
        contains((p(Str | I2 | Cap)) | MRest) /\
       (MRest =/= emptyMarking) /\
        equalMarking(MNet =?= MRest) .
ceq equalMarking(
    (p(Str | I | Cap) , MNet) =?= (p(Str | I | Cap) , MRest))
    = true
    if not(contains((p(Str | I | Cap)) | MNet)) /\
        not(contains((p(Str | I | Cap)) | MRest)) /\
       (MRest =/= emptyMarking) /\
        equalMarking(MNet =?= MRest) .
ceq equalMarking(
    (p(Str | I1 | Cap) , MNet) =?= (p(Str | I2 | Cap))) = true
    if not(contains((p(Str | I2 | Cap)) | MNet)) .
eq equalMarking(
   (PNet) =?= (PRule))
   = false [owise] .

*** READING: PLACE, PRE, POST
op emptyNeighbourForPlace(_, _, _) : Places Pre Post -> Bool .
eq emptyNeighbourForPlace(P, 
     pre{ (T --> P , PRest) , MTupleRest },
     Post) = false .
eq emptyNeighbourForPlace(P, 
     Pre,
     post{ (T --> P , PRest) , MTupleRest }) = false .
eq emptyNeighbourForPlace(P, Pre, Post) = true [owise] .
\end{lstlisting}

\subsection{Multi-Set for used Identifiers}\label{sec:identifierKrams}
One problem of Maude is the missing garbage collection\footnote{\url{http://maude.cs.uiuc.edu/maude1/manual/maude-manual-html/maude-manual_42.html}, retrieval on 20/10/2014}. This can result in an overflow if a rule inserts a node (place or transition), because each new node gets an identifier.\\
To solve this problem a multi-set of unique identifiers is used. It requires a modification of the \textit{Configuration} definition which is introduced in \autoref{lst:RuleAndConfiguration}. Now it contains an integer for the \textit{maxID} and for the defined \textit{step\_size}. Further, it has two sets for the place and transition identifiers. The implementation can be found in \autoref{lst:newConfiguration}.

\begin{minipage}[H]{0.95\linewidth}
\centering
\begin{lstlisting}[breaklines=true, language=maude, caption={Extending the \textit{Configuration} with the identifier multi-set}, label={lst:newConfiguration}, mathescape=true]
*** READING: NET SET<RULE> MAXID STEP_SIZE PID TID
op ______ : 
   Net Rule Int Int IDPoolPlace IDPoolTransition 
   ->
   Configuration .
\end{lstlisting}
\end{minipage}

\noindent The usage of the defined fields in \autoref{lst:newConfiguration} are useful when a rule deletes or adds a node as a place or a transition. An example of the implementation for a rule is shown in \autoref{lst:idMultiSetRuleImpl}. The transition \textit{t("X" $|$ Irule1031)} has to be deleted here. The identifier of this node is contained in the left side of the rule. The variable \textit{Irule1031} holds the current value which is reused in line 21, where the ID is added into the new identifier multi-set \textit{AidTRest1}. The next step uses this multi-set to receive a new identifier for the new transition with the identifier \textit{AidT1}. The getter-operator in line 22 sets the value for the new transition. Further, the new identifier must be deleted from the old identifier multi-set. The last step sets the \textit{maxID} to its new value, if the \textit{max-step} is overrun.

\begin{minipage}[H]{0.95\linewidth}
\centering
\begin{lstlisting}[breaklines=true, language=maude, caption={Save old identifier and receive a new from the identifier multi-sets}, label={lst:idMultiSetRuleImpl}, mathescape=true]
crl [R1-PNML] :
     ...
     transitions{ t("X" | $\textbf{Irule1031}$) : TRest } ,
     ...
     MaxID
     StepSize
     aidPlace{ AidPRest }
     aidTransition{ AidTRest }
     =>
     ...
     transitions{ t("X" | $\textbf{AidT1}$) : TRest } ,
     pre{ (t("X" | $\textbf{AidT1}$) --> p("A" | Irule1013 | 2147483647)) , 
            MTupleRest1 } ,
     post{ (t("X" | $\textbf{AidT1}$) --> p("A" | Irule1016 | 2147483647)) , 
            MTupleRest2 } ,
     ...
     NewMaxID
     StepSize
     aidPlace{ AidPRest }
     aidTransition{ AidTRest2 }
     if AidTRest1 := addOldID(AidTRest | $\textbf{Irule1031}$) /\
         $\textbf{AidT1}$ := getAid(AidTRest1 | MaxID | StepSize) /\
         AidTRest2 := removeFirstElement(AidTRest | MaxID | StepSize) /\
         NewMaxID := correctMaxID(MaxID | StepSize | 1) .
\end{lstlisting}
\end{minipage}

\noindent Every operator which is used in the \autoref{lst:idMultiSetRuleImpl} is shown in \autoref{lst:idMultiSet}. This listing contains the implementation of the fill-operator which generates new identifiers if the set is empty. Further, it has operators which can be used to receive or set an identifier to a multi-set of identifiers. It is not necessary to differ between places and transitions because the operators can be generic programmed. Each operator has parameters which can take both multi-sets. So the \textit{getAid}-operator provides the function to receive the first element of a multi-set. Otherwise, the \textit{removeFirstElement}-operator can delete this first element of a given multi-set. Moreover, the \textit{addOldID}-operator adds an element into a multi-set and the \textit{correctMaxID}-operator defines a new \textit{maxID} if it is necessary.

\begin{minipage}[H]{0.95\linewidth}
\centering
\begin{lstlisting}[breaklines=true, language=maude, caption={Identifier multiset implementation}, label={lst:idMultiSet}, mathescape=true]
*** READING: IDSET MAXID COUNTER INTERNAL-VAR
op fill(_|_|_|_) : Int Int Int Int -> Int .
eq fill(I | MaxID | 0 | Count) = I . 
ceq fill(IRest | MaxID | Count | I)
     = fill((MaxID + I, (IRest)) | MaxID | (Count - 1) | (I - 1)) 
     if I >= Count .
eq fill(I1 | MaxID | I2 | Count ) = I1 [owise] .

*** READING: CURRENT_SET MAXID STEP_SIZE
op getAid(_|_|_) : Int Int Int -> Int .
ceq getAid(I1 , (IRest) | MaxID | StepSize) = I1 if I1 =/= emptyIDSet .
eq getAid(SetOfInts | MaxID | StepSize) 
    = getAid(fill(SetOfInts | MaxID | StepSize | StepSize) 
              | MaxID + MaxID | StepSize) [owise] .

*** READING: CURRENT_SET MAXID STEP_SIZE
op removeFirstElement(_|_|_) : Int Int Int -> Int .
eq removeFirstElement(emptyIDSet | MaxID | StepSize) = 
                          fill(emptyIDSet | MaxID | StepSize | StepSize) .
ceq removeFirstElement(I1 , (IRest) | MaxID | StepSize) = IRest 
                           if I1 =/= emptyIDSet [owise] .
*** READING: CURRENT_SET OLD_ID
op addOldID(_|_) : Int Int -> Int .
eq addOldID(SetOfInts | I) = I, (SetOfInts) .

*** READING: MAXID STEP_SIZE NEW_ID_COUNT
op correctMaxID(_|_|_) : Int Int Int -> Int .
ceq correctMaxID(MaxID | StepSize | Count) 
                    = correctMaxID(MaxID + StepSize | StepSize | 
                    Count - StepSize) 
                    if Count > StepSize .
eq correctMaxID(MaxID | StepSize | Count) = MaxID .
\end{lstlisting}
\end{minipage}

\section{Transformation}
This section includes the architecture of the conversion process as well as the results of some model checking formula.

\subsection{Architecture}
\begin{figure}[H]
\centering
\includegraphics[keepaspectratio=true, width=\linewidth]{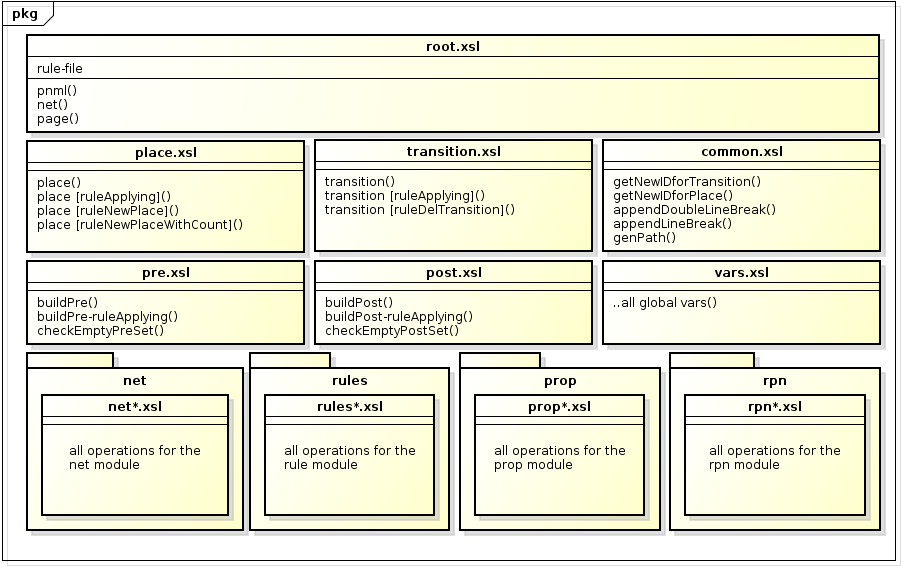}
\caption{Structure of the stylesheets for the conversion}
\label{fig:stylesheetStructure}
\end{figure}

The output base of \textit{ReConNet} consists an extension of PNML\footnote{\url{http://www.pnml.org/}, retrieval on 23/09/2014}. PNML is a XML-based standard for the Petri net export. The graphical editor \textit{ReConNet} uses this standard for the persistence of developed nets. In addition to the pure PNML-standard, a rule is stored with its three net in a PNML file.\\
Based on PNML, this work uses XSL to realise the conversion. The result uses the Maude modules which are defined before the conversion.The sorts for \textit{Places, Transitions} and the \textit{net} itself are previously defined. And further it contains the logic of firing or the identification of the dangling condition (see also the definition of all modules in the listings above).\\
The XSL process is designed with the separation of the global types as \textit{places, transitions, pre} or \textit{post}. Further, it has the specific sub xsl-templates for the conversions as in the \textit{net, rules, prop} or \textit{rpn}. The structure is summarized in \autoref{fig:stylesheetStructure}. The global types are defined above the specific modules that are grouped together in separate packages.

\noindent The difference to the first project is that this approach is independent of \textit{ReConNets} implementation. The first project uses the \textit{persistence}-module of \textit{ReConNets} to load PNML-files (net or rule) \citep{project1}. This approach is superior because it is built up directly on the PNML data. The conversion process is written with XSL, which provides a well known language. The interface between this approach and \textit{ReConNet} is realised through an export to the PNML files that can be used with a net or few rules.

\subsection{Results}
Based on the result of the conversion process it is possible to use Maude's LTL implementation. The \textit{prop.maude}-module includes all the necessary code for the LTL process (see \autoref{lst:subsortOfState}). It subsorts the \textit{Configuration}-typ under \textit{state}, which is required (for the \textit{Configuration}-typ see \autoref{lst:RuleAndConfiguration}).

\begin{minipage}[H]{0.95\linewidth}
\centering
\begin{lstlisting}[breaklines=true, language=maude, caption={Sub sorting of \textit{Configuration} with \textit{State}}, label={lst:subsortOfState}, mathescape=true]
including SATISFACTION .
subsort Configuration < State .
\end{lstlisting}
\end{minipage}

\noindent In the first example, the deadlock freedom of example \autoref{fig:petrinetz} will be shown. The formula is based on the box- and diamond-operator. In total it describes the semantic of liveness. It means that a property is globally (box) repeatedly (diamond) true. To write the liveness property for the reconfigurable Petri net modules, the following line can be used:

\[\text{rew modelCheck}(\text{initial},\text{ }[]<>\text{ enabled})\text{ }.\label{form:liveness}\]

\noindent It uses the Maude \textit{modelCheck}-operator with the initial configuration (net, marking, rules and global variables) and the formula \textit{[]$<>$ enabled}. The formula is based on the three operators $[]$ and $<>$ as well as the \textit{enabled}-operator (for the definition see \autoref{lst:LTLProp}).\\
In terms of this example the following output in \autoref{lst:enabledOfN1} results, if the trace is enabled.

\begin{lstlisting}[breaklines=true, language=maude, caption={Counterexample of a deadlock}, label={lst:enabledOfN1}, mathescape=true]
Maude> rew modelCheck(initial, []<> enabled ) .
rewrite in NET : modelCheck(initial, []<> enabled) .
rewrites: 197 in 0ms cpu (0ms real) (270604 rewrites/second)
result ModelCheckResult: counterexample(
...
{net(places{p("A" | 2 | 2147483647),p("A" | 3 | 2147483647),
            p("A" | 4 | 2147483647)},
     transitions{t("T" | 6) : t("T" | 7) : t("T" | 26)},
     pre{(t("T" | 6) --> p("A" | 2 | 2147483647)),
         (t("T" | 7) --> p("A" | 3 | 2147483647)),
          t("T" | 26) --> p("A" | 3 | 2147483647)},
     post{(t("T" | 6) --> p("A" | 4 | 2147483647)),
          (t("T" | 7) --> p("A" | 2 | 2147483647)),
           t("T" | 26) --> p("A" | 4 | 2147483647)},
     marking{p("A" | 4 | 2147483647) ; p("A" | 4 | 2147483647)})
 rule(l(net(places{p("A" | 17 | 2147483647),
                      p("A" | 20 | 2147483647)},
            transitions{t("T" | 24)},
            pre{t("T" | 24) --> p("A" | 17 | 2147483647)},
            post{t("T" | 24) --> p("A" | 20 | 2147483647)},
            marking{p("A" | 17 | 2147483647)})),
      r(net(places{p("A" | 17 | 2147483647),p("A" | 20 | 2147483647)},
            transitions{t("T" | 26)},
            pre{t("T" | 26) --> p("A" | 20 | 2147483647)},
            post{t("T" | 26) --> p("A" | 17 | 2147483647)},
            marking{p("A" | 17 | 2147483647)})))
 26
 10
 aidPlace{26,(27,(28,(29,(30,(31,(32,(33,(34,(35,(
 36))))))))))}
 aidTransition{26,(27,(28,(29,(30,(31,(32,(33,(34,(35,(36))))))))))}
 ,deadlock})
\end{lstlisting}

\begin{figure}[H]
\centering
\begin{tikzpicture}[node distance=1.3cm,>=stealth',bend angle=45,auto]
  \tikzstyle{place}=[circle,thick,draw=blue!75,fill=blue!20,minimum size=6mm]
  \tikzstyle{transition}=[rectangle,thick,draw=black!75,
  			  fill=black!20,minimum size=6mm]

  \tikzstyle{every label}=[black]

  \begin{scope}
    \node [place] (A1) [label=above:$A_{3}$] {};

    \node [transition] (T1) [right of=A1, label=above:$T_{26}$] {}
      edge [pre, above] (A1);

    \node [place,tokens=2] (A2) [xshift=8mm, below of=T1, label=above:$A_{4}$] {}
      edge [pre, above] (T1);

    \node [transition] (T2) [xshift=-8mm, below of=A2, label=above:$T_{6}$] {}
      edge [post, above] (A2);

    \node [place] (A3) [left of=T2, label=above:$A_{2}$] {}
      edge [post, above] (T2);

    \node [transition] (T3) [xshift=-8mm, above of=A3, label=above:$T_{7}$] {}
      edge [pre, above] (A1)
      edge [post, above] (A3);
  \end{scope}
\end{tikzpicture}
\caption{State of $N_{1}$ with a deadlock ($r_{1}$ can not be applied)}
\label{fig:deadlockState}
\end{figure}
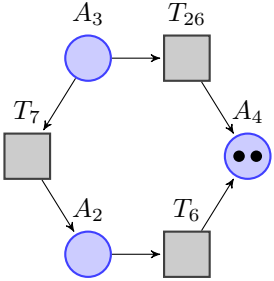

\noindent The meaning of this counterexample is that the rule consists of this marking. It is possible that all tokens are on one place. Furthermore, this place has only incoming arcs which results in a deadlock. The net-state with this deadlock is modelled in \autoref{fig:deadlockState}.

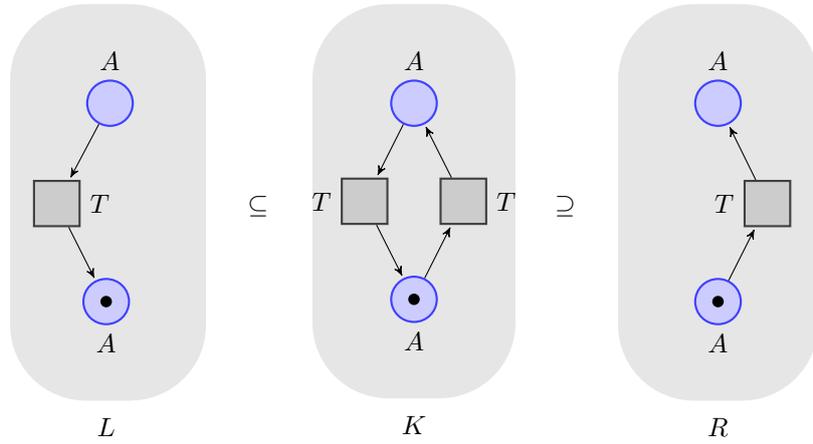
\begin{figure}[H]
\centering
\begin{tikzpicture}[node distance=1.3cm,>=stealth',bend angle=45,auto]
  \tikzstyle{place}=[circle,thick,draw=blue!75,fill=blue!20,minimum size=6mm]
  \tikzstyle{transition}=[rectangle,thick,draw=black!75,
  			  fill=black!20,minimum size=6mm]

  \tikzstyle{every label}=[black]

  \begin{scope}
    \node [place] (A1_l) 
      [label=above:$A$] {};
	    
    \node [transition] (T_l) 
      [yshift=-1.33cm, xshift=-20mm, right of=A1_l, label=right:$T$] {}
      edge [pre, right] (A1_l);	  
	  
    \node [xshift=1.35cm, yshift=0.35cm, right of=T_l, label=below:$\subseteq$] {}; 
	    
	\node [place,tokens=1] (A2_l) 
	  [xshift=6.5mm, below of=T_l, label=below:$A$] {}
      edge [pre, right] (T_l);
	
	\node [below of=A2_l, label=below:$L$] {};
  \end{scope}
  
  \begin{scope}[xshift=4cm]
    \node [place] (A1_k) 
      [label=above:$A$] {};

    \node [transition] (T1_k) 
      [xshift=-6.5mm, below of=A1_k, label=left:$T$] {}
      edge [pre, left] (A1_k);
      
    \node [transition] (T2_k) 
      [right of=T1_k, label=right:$T$] {}
      edge [post, right] (A1_k);

    \node [place,tokens=1] (A2_k) 
      [xshift=6.5mm, below of=T1_k, label=below:$A$] {}
      edge [post, left] (T2_k)
      edge [pre, left] (T1_k);

	\node [below of=A2_k, label=below:$K$] {};
  \end{scope}

  \begin{scope}[xshift=8cm]
    \node [place] (A1_r) 
      [label=above:$A$] {};

    \node [transition] (T_r) 
      [yshift=-1.33cm, xshift=-6.5mm, right of=A1_r, label=left:$T$] {}
      edge [post, right] (A1_r);	  

    \node [yshift=0.35cm, xshift=-1.36cm, left of=T_r, label=below:$\supseteq$] {};  

	\node [place,tokens=1] (A2_r) 
	  [xshift=-6.5mm, below of=T_r, label=below:$A$] {}
      edge [post, right] (T_r);

	\node [below of=A2_r, label=below:$R$] {};
  \end{scope}

  \begin{pgfonlayer}{background}
    \filldraw [line width=20mm, join=round, black!10]
      (A1_l.north  -| A2_l.east)  rectangle
      (A2_l.south  -| A1_l.west);
    \filldraw [line width=20mm, join=round, black!10]
      (A1_k.north  -| T1_k.east)  rectangle
      (A2_k.south  -| T2_k.west);
    \filldraw [line width=20mm, join=round, black!10]
      (A1_r.north  -| A2_r.east)  rectangle
      (A2_r.south  -| A1_r.west);
  \end{pgfonlayer}
\end{tikzpicture}
\caption{Example rule $r2$ which changes the direction of the arc (different marking in contrast to $r_{1}$ in \autoref{fig:exampleRule})}
\label{fig:exampleRuleNo2}
\end{figure}

\noindent In assumption that the markings were changed on the places within the rule (see \autoref{fig:exampleRuleNo2}) the result is varied, shown in \autoref{lst:enabledOfN1WithChangedRule}. It shows, that the net \textit{N1} and rule \textit{r2} are deadlock free. The new rule prevents the situation in \autoref{fig:deadlockState}, where the marking can be located on one place which has only incoming arcs. Hence, two situations are possible. First, the rule is not enabled when a marking is at a place, where all arcs are starting. The net itself can fire. 
On the other hand, a marking will be placed on a place where one or more arcs are incoming. For this case, the rule can be used. The result is that a transition is enabled now and will also continue to be used for the token game. In either situation, an operator of \autoref{lst:LTLProp} is enabled since the \textit{enabled}-operator is defined for the firing and transformation step.

\noindent Now the result of the formula is true and Maude prints some information such as the rewrite count.

\begin{minipage}[H]{0.95\linewidth}
\centering
\begin{lstlisting}[breaklines=true, language=maude, caption={Counterexample for the deadlock freeness}, label={lst:enabledOfN1WithChangedRule}, mathescape=true]
Maude> rew modelCheck(initial, []<> enabled ) .
rewrite in NET : modelCheck(initial, []<> enabled) .
Debug(1)> rew [1] modelCheck(initial, []<> enabled ) .
rewrite [1] in NET : modelCheck(initial, []<> enabled) .
rewrites: 6575268 in 20240ms cpu (20243ms real) (324865 rewrites/second)
result Bool: true
\end{lstlisting}
\end{minipage}

\noindent In addition, the enable-test can only be realised for the transitions with the \textit{t-enabled}-operator. It only consists of the transitions and no rules (see \autoref{lst:tenabled}).

\begin{minipage}[H]{0.95\linewidth}
\centering
\begin{lstlisting}[breaklines=true, language=maude, caption={Counterexample for a t-enabled test}, label={lst:tenabled}, mathescape=true]
Maude> rewrite [1] in NET : modelCheck(initial, []<> t-enabled) .
rewrites: 1521 in 4ms cpu (6ms real) (380250 rewrites/second)
result Bool: true
\end{lstlisting}
\end{minipage}

\noindent The \textit{reachable}-operator is designed to find a given marking. The formal definition of this operator is:

\[ \text{op reachable} : \text{Markings} \rightarrow \text{Prop} . \]

\noindent Therefore, it is possible to test one or many tokens. In the example bellow (see \autoref{lst:reachableOfaMarkingInN1}) two tokens ($A_{3}$ and $A_{3}$) are searched in $N_{1}$. Maude's result describes a situation where it found a deadlock with two tokens on $A_{2}$.

\begin{minipage}[H]{0.95\linewidth}
\centering
\begin{lstlisting}[breaklines=true, language=maude, caption={Counterexample for a reachable test}, label={lst:reachableOfaMarkingInN1}, mathescape=true]
Maude> rew modelCheck(initial, <> reachable(p("A" | 3 | 2147483647) ; p("A" | 3 | 2147483647))) .
rewrite in NET : modelCheck(initial, <> reachable(p("A" | 3 | 2147483647) ; p("A" | 3 | 2147483647))) .
rewrites: 141 in 0ms cpu (0ms real) (262081 rewrites/second)
result ModelCheckResult: counterexample(
...
{net(places{p("A" | 2 | 2147483647), p("A" | 3 | 2147483647),
              p("A" | 4 | 2147483647)},
     transitions{t("T" | 5) : t("T" | 7) : t("T" | 26)},
     pre{(t("T" | 5) --> p("A" | 4 | 2147483647)),
          (t("T" | 7) --> p("A" | 3 | 2147483647)),
           t("T" | 26) --> p("A" | 4 | 2147483647)}, 
     post{(t("T" | 5) --> p("A" | 3 | 2147483647)),
           (t("T" | 7) --> p("A" | 2 | 2147483647)),
            t("T" | 26) --> p("A" | 2 | 2147483647)},
     marking{p("A" | 2 | 2147483647) ; 
              p("A" | 2 | 2147483647)}) 
 rule(l(net(places{p("A" | 17 | 2147483647), p("A" | 20 | 2147483647)}, 
              transitions{t("T" | 24)}, 
              pre{t("T" | 24) --> p("A" | 17 | 2147483647)},
              post{t("T" | 24) --> p("A" | 20 | 2147483647)}, 
              marking{p("A" | 17 | 2147483647)})),
      r(net(places{p("A" | 17 | 2147483647),p("A" | 20 | 2147483647)}, 
              transitions{t("T" | 26)}, 
              pre{t("T" | 26) --> p("A" | 20 | 2147483647)},
              post{t("T" | 26) --> p("A" | 17 | 2147483647)}, 
              marking{p("A" | 17 | 2147483647)})))
 26
 10
 aidPlace{26,(27,(28,(29,(30,(31,(32,(33,(34,(35,(36))))))))))} 
 aidTransition{26,(27,(28,(29,(30,(31,(32,(33,(34,(35,(36))))))))))}
 ,deadlock}
\end{lstlisting}
\end{minipage}

\noindent An example where the LTL-formal is true can be found in \autoref{lst:reachableOftheInitMarking}. This is a basic test where the formula verifies the reachability of the initial marking. Maude reached true after six steps because the marking already contains the search parameter.

\begin{minipage}[H]{0.95\linewidth}
\centering
\begin{lstlisting}[breaklines=true, language=maude, caption={Counterexample for a reachable of the initial marking}, label={lst:reachableOftheInitMarking}, mathescape=true]
Maude> rew modelCheck(initial, <> reachable(p("A" | 3 | 2147483647) ; p("A" | 4 | 2147483647))) .
rewrite in NET : modelCheck(initial, <> reachable(p("A" | 3 | 2147483647) ; p("A" | 4 | 2147483647))) .
rewrites: 6 in 0ms cpu (0ms real) (50000 rewrites/second)
result Bool: true
\end{lstlisting}
\end{minipage}

\noindent A more complex example can be found in \autoref{lst:reachableOfAMarking}. The example verified that two markings ($A_{4}$ and $A_{4}$) are reachable from the initial marking. The formula differs from the formula in the example above (see \autoref{lst:reachableOftheInitMarking}). It negated the formula, so that the result contains a path to this marking, or otherwise it returns true.

\begin{lstlisting}[breaklines=true, language=maude, caption={Counterexample for a reachable of the initial marking}, label={lst:reachableOfAMarking}, mathescape=true]
Maude> rew [1] modelCheck(initial, ~( <> reachable(p("A" | 4 | 2147483647) ; p("A" | 4 | 2147483647)))) .
rewrite [1] in NET : modelCheck(initial, ~ <> reachable(p("A" | 4 | 2147483647) ; p("A" | 4 | 2147483647))) .
rewrites: 174 in 0ms cpu (0ms real) (~ rewrites/second)
result ModelCheckResult: counterexample(
...
{ net(places{p("A" | 2 | 2147483647),p("A" | 3 | 2147483647),p("A" | 4 | 2147483647)}, 
  transitions{t(
    "T" | 5) : t("T" | 6) : t("T" | 7)}, 
  pre{(t("T" | 5) --> p("A" | 4 | 2147483647)),
       (t("T" | 6) --> p("A" | 2 | 2147483647)),
        t("T" | 7) --> p("A" | 3 | 2147483647)}, 
  post{(t("T" | 5) --> p("A" | 3 | 2147483647)),
        (t("T" | 6) --> p("A" | 4 | 2147483647)),
         t("T" | 7) --> p("A" | 2 | 2147483647)},
  $\textbf{marking{p("A" | 4 | 2147483647) ; p("A" | 4 | 2147483647)})}$ 
  rule(
  l(net(
      places{p("A" | 17 | 2147483647),p("A" | 20 | 2147483647)}, 
      transitions{t("T" | 24)}, 
      pre{t("T" | 24) --> p("A" | 17 |
    2147483647)}, 
      post{t("T" | 24) --> p("A" | 20 | 2147483647)}, 
      marking{p("A" | 17 | 2147483647)})),
  r(net(
      places{p("A" | 17 | 2147483647),p("A" | 20 | 2147483647)}, 
      transitions{t("T" | 26)}, 
      pre{t("T" | 26) --> p("A" | 20 | 2147483647)}, 
      post{t("T" | 26) --> p("A" | 17 | 2147483647)}, 
      marking{p(
    "A" | 17 | 2147483647)}))) 
  26,{'fire} }, nil)
\end{lstlisting}

\noindent Based on the \textit{reachable}-operator it is possible to describe the liveness-condition, which means that a marking can be reached again. This includes all possible firing steps or rule applying.\\
In the example net $N_{1}$ the test will return a situation where the net arrests in a deadlock. All tokens are placed on $A_{4}$ which has only incoming arcs. Hence, no rule (in this example $r_{1}$) can be applied (for a graphical representation of the result see also \autoref{fig:deadlockState}).

\begin{minipage}[H]{0.95\linewidth}
\centering
\begin{lstlisting}[breaklines=true, language=maude, caption={Counterexample for the liveness-condition for $A_{3}$}, label={lst:reachableOftheInitMarking}, mathescape=true]
Maude> rew modelCheck(initial, []<> reachable(p("A" | 3 | 2147483647))) .
rewrite in NET : modelCheck(initial, []<> reachable(p("A" | 3 | 2147483647))) .
rewrites: 197 in 0ms cpu (0ms real) (285094 rewrites/second)
result ModelCheckResult: counterexample(
...
{net(places{p("A" | 2 | 2147483647), p("A" | 3 | 2147483647),
              p("A" | 4 | 2147483647)}, 
     transitions{t("T" | 6) : t("T" | 7) : t("T" | 26)}, 
     pre{(t("T" | 6) --> p("A" | 2 | 2147483647)), 
          (t("T" | 7) --> p("A" | 3 | 2147483647)),
           t("T" | 26) --> p("A" | 3 | 2147483647)},
     post{(t("T" | 6) --> p("A" | 4 | 2147483647)),
           (t("T" | 7) --> p("A" | 2 | 2147483647)),
            t("T" | 26) --> p("A" | 4 | 2147483647)},
     marking{p("A" | 4 | 2147483647) ; p("A" | 4 | 2147483647)})
 rule(l(net(places{p("A" | 17 | 2147483647),p("A" | 20 | 2147483647)}, 
              transitions{t("T" | 24)}, 
              pre{t("T" | 24) --> p("A" | 17 | 2147483647)},
              post{t("T" | 24) --> p("A" | 20 | 2147483647)},
              marking{p("A" | 17 | 2147483647)})),
      r(net(places{p("A" | 17 | 2147483647),p("A" | 20 | 2147483647)}, 
             transitions{t("T" | 26)}, 
             pre{t("T" | 26) --> p("A" | 20 | 2147483647)},
             post{t("T" | 26) --> p("A" | 17 | 2147483647)}, 
             marking{p("A" | 17 | 2147483647)}))),deadlock})
\end{lstlisting}
\end{minipage}

\subsection{Tests}
Test cases are separated into two steps for the conversion to Maude and a verification with Maude's LTL model checker. XSLT provides a test mechanism with XSLTunit\footnote{\url{http://xsltunit.org/}, retrieval on 20/10/2014}, which is implemented for the used XSL processors SAXON\footnote{\url{http://saxon.sourceforge.net/}, retrieval on 20/10/2014}. Further, the Maude test cases are realised with JUnit which implements a Java process. It is executing a shell script including a set of commands such as verifications for a deadlock. The \textit{Maude Development Tools}\footnote{\url{http://moment.dsic.upv.es/mdt/}, retrieval on 20/10/2014} can not be used, since there is no support for model checking commands.

\section{Evaluation}
This section presents a first step of the performance evaluation of this approach. The evaluation is based on two steps. At first, a net is converted into the Maude modules. And after that step it is tested with the liveness formula (see the first formula in section \autoref{form:liveness}).\\
The conversion uses a net which is build as a circle and the rule $r_{1}$ (see \autoref{fig:exampleRule}). Further, it contains one token at place $P_{1}$. The structure connects a place with two transitions (one for the pre and vice versa). Hence, it is possible to build a test which shows the performance of a net which can be scaled with the size of nodes (places and transitions).\\
For this work four net sizes are used, which allow to make a meaningful statement. Each net has the same semantic and should return true. Hence, only the runtime meta-data such as rewrite count and time are different. The conversion process runs in each case with nearly the same time (see \autoref{tab:transformationTime}). Further, the rewrites are grown linear with the size of nodes. It takes 109 rewrites for 10 places and transitions. If the net has twice as many nodes as in the first example, it takes 209 rewrites. Only the used time grows exponentially. It changes from 13 ms to 23994 ms. If the net receives only 4 new nodes, it grows to 189939 ms (see \autoref{tab:runtime}). The resulting state-space explosion was expectable as a well known issue of LTL \citep{valmari1998state}.\\
All tests are realised on a Thinkpad X230 with an Intel® Core™ i5-3320M CPU with 4 cores (2.60GHz) and 16 GB RAM. It was implemented on a Ubuntu 12.04 which is build up the 3.14.17-031417-generic kernel.

\begin{table}[H]
\parbox{.45\linewidth}{
\centering
\begin{tabular}{|l|c|r|}
\hline \multicolumn{1}{|c|}{p x t} & \multicolumn{1}{c|}{} & \multicolumn{1}{|c|}{time [ms]} \\
\hline 10x10 & in & 1701 \\ 
\hline 20x20 & in & 1737 \\ 
\hline 22x22 & in & 1704 \\ 
\hline 23x23 & in & 1737 \\ 
\hline 
\end{tabular}
\caption{Conversion from PNML to Maude}
\label{tab:transformationTime}
}
\hfill
\parbox{.45\linewidth}{
\centering
\begin{tabular}{|l|c|r|}
\hline \multicolumn{1}{|c|}{p x t} & \multicolumn{1}{c|}{rew} & \multicolumn{1}{|c|}{time [ms]} \\
\hline 10x10 & 109 in & 13 \\ 
\hline 20x20 & 209 in & 23994 \\ 
\hline 22x22 & 229 in & 189939 \\ 
\hline 23x23 & 239 in & 834358 \\ 
\hline 
\end{tabular}
\caption{Verification of liveness}
\label{tab:runtime}
}
\end{table}

\section{Future Work}
An outlook of the following work is shown in this section. The formal founding represents the essential part of the conversion from a reconfigurable Petri net to Maude. The aim is the verification of the soundness and correctness for the conversion and target representation of the net in Maude.\\
Furthermore, an integration of parts from a reconfigurable Petri net such as negative application conditions (NACs) (see \citep{rein2008negative}) or decorations (see \citep{ede2012reconnet}) should be realised. This enables the verification of the nets and rules from the \textit{Living Place Hamburg} \citep{ede2012reconnet}.\\
Finally, a benchmarking is necessary between this approach and a tool such as Charlie\footnote{\url{http://www-dssz.informatik.tu-cottbus.de/DSSZ/Software/Charlie}, retrieval on 19/10/2014} to obtain a meaningful statement. This implies a way which converts a reconfigurable Petri net into a net that can be used from other verification tools. The main challenge is to realise a conversion which contains the net and all possible rule conditions.

\section{Conclusion}
This paper presents an approach which enables LTL model checking for reconfigurable Petri nets. The intention was to use Maude with the equation- and rewrite logic for the transformation result. Maude itself includes modules for an on-the-fly model checking, based on the state defined by the new modules of this work.\\
The resulting modules can be used for the modelling as the formal definition of the reconfigurable Petri nets. This means that the modules include a structure of sorts and with associated operators which allow a tool or user to write a clear formal definition of the net as well as a set of rules. Furthermore, conditions as for example the dangling condition are also included by the modules defined by this paper.\\
Finally, the evaluation shows that the defined modules have problems with the size of the net. One problem is that a rule can add a place or transition. Hence, it is necessary to get new identifiers for this elements. If a rule inserts a new node and this new node receives a new identifier, the problem results in an infinite behaviour. Currently this problem is not solved. Unless it exists a countable reuse of identifiers. For a example it is possible that a rule deletes one element and inserts a new one. The old unused identifier can be recycled. For this special formula the model checking process returns. If the rule creates infinite nodes (each use inserts a new transition) it has no chance to receive a result for this formula.

\newpage
\section{Summary}
The aim of this work is to allow LTL for reconfigurable Petri nets. The tool \textit{ReConNet} is the base, which makes it possible to create a reconfigurable Petri net. Further, it includes a possibility to export a net and a set of rules as PNML-files. This files are the origin for this approach.\\
The approach realises Maude modules which can be used in a LTL process. Maude includes a module for an on-the-fly model checking process. The new modules consist of 4 separated parts. First a module contains the definition for an algebraic reconfigurable Petri net. The aim of the module is to allow a writing of a net and a set of rules as it is provided by the mathematical notation. Furthermore, it supports an activation and firing of a net. The next module contains the rule definitions. Rewrite rules are used to design a rule which uses the pattern matching of Maude for the possibility to use this rule. Each rule ensures that the dangling condition is maintained. Further, an identifier multi-set is used to cache unused identifier for places and transitions. This caching allows the process to verify a formula, if the number of identifiers is limited. Next, a module contains the definition for operators which are necessary for the LTL formulae. For example the \textit{enabled}-operator can be used for the liveness condition. At last, a module includes the initial definition of a net and a set of rules. The initial state embodies the initial marking, places and transitions.\\
Based on each module LTL formulae can be verified with Maude. It returns true, or a \textit{counterExample} with an example for an error case of this formula.

\newpage
\printglossaries
\listoffigures
\newpage
\lstlistoflistings
\newpage

\bibliographystyle{natdin} 
\bibliography{Verzeichnis}

\newpage
\begin{appendices}

\section*{Reconfigurable Petri Nets - RPN.maude}

\begin{lstlisting}[language=maude]
mod RPN is
  protecting INT .
  protecting STRING .

  *** ######################################################################
  *** Local VARs
  *** ######################################################################

  var Str Str1 Str2 : String .
  var I I1 I2 I3 Cap Cap1 Cap2 Counter MaxID StepSize : Int .
  var P PRest PSet PSetL PSetR MTupleValue PreValue PostValue P1 P2 : Places .
  var T TRest TSet TSetL TSetR : Transitions .
  var Pre PreL PreR : Pre .
  var Post PostL PostR : Post .
  var MTupleRest MTupleRest1 MTupleRest2 : MappingTuple .
  var M M1 M2 MRest MRest1 MRest2 MFollow : Markings .
  var R RRest Rules : Rule .
  var aidP : IDPoolPlace .
  var aidT : IDPoolTransition .


  *** ######################################################################
  *** Petrinet N = (P, T, Pre, Post, M_0)
  *** ######################################################################

  sort Net .
  sort Places .
  sort Transitions .
  sort Pre .
  sort Post .
  sort MappingTuple .
  sort Markings .

  subsort Places < Markings .

  op emptyPlace : -> Places .
  op emptyTransition : -> Transitions .
  op emptyMappingTuple : -> MappingTuple .
  op emptyMarking : -> Markings .

  op _,_ : Places Places -> Places [ctor assoc comm id: emptyPlace] .
  op _+_ : Places Places -> Places [ctor assoc comm id: emptyPlace] .
  op _:_ : Transitions Transitions -> 
            Transitions [ctor assoc comm id: emptyTransition] .
  op _,_ : MappingTuple MappingTuple -> 
            MappingTuple [ctor assoc comm id: emptyMappingTuple] .
  op _;_ : Markings Markings -> Markings [ctor assoc comm id: emptyMarking] .

  *** READING: Pname | ID | Cap
  op p(_|_|_) : String Int Int -> Places .
  op t(_|_) : String Int -> Transitions .

  op (_-->_) : Transitions Places -> MappingTuple .

  op places{_} : Places -> Places .
  op transitions{_} : Transitions -> Transitions .
  op pre{_} : MappingTuple -> Pre .
  op post{_} : MappingTuple -> Post .
  op marking{_} : Markings -> Markings .

  *** Petrinet-tuple
  op net : Places Transitions Pre Post Markings -> Net .


  *** ######################################################################
  *** Firing of N = (P, T, Pre, Post, M_0)
  *** ######################################################################

  op calc : Markings -> Markings .
  op _plus_ : Markings Markings -> Markings .
  op _minus_ : Markings Markings -> Markings .

  *** #################################
  *** Enable AND Calc
  *** #################################

  op _ leeqth _ with _ : Places Places Int -> Bool .
  op _ <=? _ : Markings Places -> Bool .

  *** Impl - lowerEqualThan ##########

  *** place multiset is empty
  ceq emptyMarking leeqth p(Str | I | Cap1) with Counter
      = true if Counter <= Cap1 .

  *** Cap-counter is too big
  eq (p(Str | I | Cap2) ; MRest) leeqth p(Str | I | Cap2) with (Cap2 + 1) 
     = false .

  *** found same place
  ceq (p(Str | I | Cap2) ; MRest) leeqth p(Str | I | Cap2) with Counter 
      = true 
      if (MRest leeqth p(Str | I | Cap2) with (Counter + 1)) .

  *** del another place
  ceq (p(Str | I | Cap1) ; MRest) leeqth p(Str2 | I2 | Cap2) with Counter 
      = true 
      if (MRest leeqth p(Str2 | I2 | (Cap2)) with Counter) .

  *** otherwise
  eq M leeqth P with I = false [owise] .

  *** Impl - smallerAsCap #############

  eq marking{ PSet } <=? emptyPlace = true .

  ceq marking{M} <=? (P , emptyPlace)
      = true
      if M leeqth P with 0 .

  ceq marking{M} <=? (P , PRest)
      = true
      if M leeqth P with 0
         /\ PRest =/= emptyPlace
         /\ marking{M} <=? PRest .

  eq M <=? P = false [owise] .

  *** Impl - fire #####################

  crl [fire-emptyPre] : 
       net(P,
           transitions{T : TRest}, 
           pre{(T --> emptyPlace), MTupleRest1}, 
           post{(T --> PostValue), MTupleRest2}, 
           marking{M}) 
       Rules
       MaxID
       StepSize
       aidP
       aidT
       =>
       net(P, 
           transitions{T : TRest}, 
           pre{(T --> emptyPlace), MTupleRest1}, 
           post{(T --> PostValue), MTupleRest2}, 
           calc(M plus PostValue))
       Rules
       MaxID
       StepSize
       aidP
       aidT
       if calc(M plus PostValue) <=? PostValue .

  crl [fire] : 
       net(P,
           transitions{T : TRest}, 
           pre{(T --> PreValue), MTupleRest1}, 
           post{(T --> PostValue), MTupleRest2}, 
           marking{PreValue ; M}) 
       Rules
       MaxID
       StepSize
       aidP
       aidT
       =>
       net(P, 
           transitions{T : TRest}, 
           pre{(T --> PreValue), MTupleRest1}, 
           post{(T --> PostValue), MTupleRest2}, 
           calc(((PreValue ; M) minus PreValue) plus PostValue))
       Rules
       MaxID
       StepSize
       aidP
       aidT
       if calc((PreValue ; M) plus PostValue) <=? PostValue .

  *** #################################
  *** Execute Calc
  *** #################################

  eq [execute-step-minusStepEmptyPlace] : 
      calc((M minus emptyPlace) plus PostValue) =
      calc(M plus PostValue) .

  eq [execute-step-minusEnd-single] : 
      calc((p(Str | I | Cap) minus (p(Str | I | Cap))) plus PostValue) =
      marking{PostValue} .

  eq [execute-step-minusStep] : 
      calc(((p(Str | I | Cap) ; MRest1) minus (p(Str | I | Cap) + MRest2)) 
           plus PostValue) =
      calc((MRest1 minus MRest2) plus PostValue) .

  eq [execute-step-plusEnd] : 
      calc(M plus PostValue) =
      marking{M ; PostValue} .

  *** ######################################################################
  *** Rule R = (l_net, r_net)
  *** ######################################################################

  sort Rule .
  sorts LeftHandSide RightHandSide .

  op emptyRule : -> Rule .

  op _|_ : Rule Rule -> Rule [ctor assoc comm id: emptyRule] .

  op l : Net -> LeftHandSide .
  op r : Net -> RightHandSide .

  op rule : LeftHandSide RightHandSide -> Rule .


  *** ######################################################################
  *** ID Pool
  *** ######################################################################

  sort IDPoolPlace IDPoolTransition .

  op emptyIDSet : -> Int .
  op _,(_) : Int Int -> Int [comm id: emptyIDSet] .

  op aidPlace{_} : Int -> IDPoolPlace .
  op aidTransition{_} : Int -> IDPoolTransition .

  *** ######################################################################
  *** Configuration
  *** ######################################################################

  sort Configuration .

  *** READING: NET SET<RULE> MAXID STEP_SIZE PID TID
  op ______ : Net Rule Int Int IDPoolPlace IDPoolTransition -> Configuration .

endm
\end{lstlisting}
\newpage

\section*{Reconfigurable Petri Nets - RULES.maude}

\begin{lstlisting}[language=maude]
mod RULES is
  including RPN .  

  var Str : String .
  vars I I1 I2 IRest IRest2 Cap Cap1 Cap2 StepSize : Int .
  vars P PRest PRest1 PRest2 PNet PRule : Places .
  vars T TRest TSet TSetL TSetR : Transitions .
  vars Pre PreL PreR : Pre .
  vars Post PostL PostR : Post .
  vars MTupleRest MTupleRest1 MTupleRest2 : MappingTuple .
  vars M M1 M2 MRest MRest1 MRest2 MNet MFollow : Markings .
  vars R RRest : Rule .
  var N : Net .

  *** ######################################################################
  *** Rule-Conditions
  *** ######################################################################

  op contains(_, _) : Places Places -> Bool .

  eq contains(p(Str | I1 | Cap), (p(Str | I2 | Cap), PRest)) = true .
  eq contains(P, PNet) = false [owise] .

  *** equalMarking returns true if:
  *** each marking of the rule is in the marking multiset of the net
  *** and the marking set has no more markings of each marking 
  ***    inside the rule set
  *** READING: NET-MARKING, RULE-MARKING
  op equalMarking(_ =?= _) : Places Places -> Bool .

  eq equalMarking(
     P =?= P
     ) = true .

  ceq equalMarking(
      (p(Str | I1 | Cap) , MNet) =?= (p(Str | I2 | Cap) , MRest)
      ) = true
      if equalMarking(MNet =?= MRest) /\
         (MRest =/= emptyMarking) .

  ceq equalMarking(
      (p(Str | I1 | Cap) , MNet) =?= (p(Str | I2 | Cap))) = true
      if not(contains((p(Str | I2 | Cap)), MNet)) .

  eq equalMarking(
     (PNet) =?= (PRule)
     ) = false [owise] .

  *** emptyNeighbourForPlace returns true if no pre/post contains this place
  *** READING: PLACE, PRE, POST
  op emptyNeighbourForPlace(_, _, _) : Places Pre Post -> Bool .

  eq emptyNeighbourForPlace(P, 
       pre{ (T --> P , PRest) , MTupleRest },
       Post) = false .

  eq emptyNeighbourForPlace(P, 
       Pre,
       post{ (T --> P , PRest) , MTupleRest }) = false .

  eq emptyNeighbourForPlace(P, Pre, Post) = true [owise] .

  *** ######################################################################
  *** RULE ID Pool getter
  *** ######################################################################

  vars MaxID NewMaxID SetOfInts Count AidPRestSet AidPRestNewSet ISet : Int .


  *** Helper, which fill the set of IDs
  *** READING: IDSET MAXID COUNTER INTERNAL-VAR

  op fill(_|_|_|_) : Int Int Int Int -> Int .
  eq fill(I | MaxID | 0 | Count) = I . 
  ceq fill(IRest | MaxID | Count | I)
       = fill((MaxID + I, (IRest)) | MaxID | (Count - 1) | (I - 1)) 
       if I >= Count .
  eq fill(I1 | MaxID | I2 | Count ) = I1 [owise] .



  *** Getter for the new ID (place or transition)
  *** READING: CURRENT_SET MAXID STEP_SIZE

  op getAid(_|_|_) : Int Int Int -> Int .

  ceq getAid(I1 , (IRest) | MaxID | StepSize) = I1 if I1 =/= emptyIDSet .
  eq getAid(SetOfInts | MaxID | StepSize) 
        = getAid(fill(SetOfInts | MaxID | StepSize | StepSize) 
                 | MaxID + MaxID | StepSize) [owise] .



  *** Remove the first element from this multiset
  *** READING: CURRENT_SET MAXID STEP_SIZE

  op removeFirstElement(_|_|_) : Int Int Int -> Int .

  eq removeFirstElement(emptyIDSet | MaxID | StepSize) = 
       fill(emptyIDSet | MaxID | StepSize | StepSize) .
  ceq removeFirstElement(I1 , (IRest) | MaxID | StepSize) = IRest 
        if I1 =/= emptyIDSet [owise] .



  *** Add unused ID
  *** READING: CURRENT_SET OLD_ID

  op addOldID(_|_) : Int Int -> Int .

  eq addOldID(SetOfInts | I) = I, (SetOfInts) .



  *** Correct the MaxID if new IDs are generated
  *** READING: MAXID STEP_SIZE NEW_ID_COUNT

  op correctMaxID(_|_|_) : Int Int Int -> Int .

  ceq correctMaxID(MaxID | StepSize | Count) 
        = correctMaxID(MaxID + StepSize | StepSize | Count - StepSize) 
        if Count > StepSize .
  eq correctMaxID(MaxID | StepSize | Count) = MaxID .

  *** ######################################################################
  *** RULE IMPLEMENTATION
  *** ######################################################################

  vars Irule103 Irule104 Irule102 Irule107 Irule105 Irule106 Irule1017 Irule1020 Irule1024 Irule2017 Irule2020 Irule2024 Irule3017 Irule3020 Irule3026 : Int .

  vars AidP AidPRest AidT1 AidT AidTRest2 AidTRest1 AidTRest : Int .

  crl [R1-PNML] :
    net(
    places{ p("A" | Irule1017 | 2147483647) , p("A" | Irule1020 | 2147483647) , PRest } ,
    transitions{ t("T" | Irule1024) : TRest } ,
    pre{ (t("T" | Irule1024) --> p("A" | Irule1017 | 2147483647)) , MTupleRest1 } ,
    post{ (t("T" | Irule1024) --> p("A" | Irule1020 | 2147483647)) , MTupleRest2 } ,
    marking{ p("A" | Irule1017 | 2147483647) ; MRest }
    )
    rule(
    l( net(
    places{ p("A" | Irule2017 | 2147483647) , p("A" | Irule2020 | 2147483647) } ,
    transitions{ t("T" | Irule2024) } ,
    pre{ (t("T" | Irule2024) --> p("A" | Irule2017 | 2147483647)) } ,
    post{ (t("T" | Irule2024) --> p("A" | Irule2020 | 2147483647)) } ,
    marking{ p("A" | Irule2017 | 2147483647) }
    ) ) ,
    r( net(
    places{ p("A" | Irule3017 | 2147483647) , p("A" | Irule3020 | 2147483647) } ,
    transitions{ t("T" | Irule3026) } ,
    pre{ (t("T" | Irule3026) --> p("A" | Irule3020 | 2147483647)) } ,
    post{ (t("T" | Irule3026) --> p("A" | Irule3017 | 2147483647)) } ,
    marking{ p("A" | Irule3017 | 2147483647) }
    ) ) )
    | RRest
    MaxID
    StepSize
    aidPlace{ AidPRest }
    aidTransition{ AidTRest }
    =>
    net(
    places{ p("A" | Irule1017 | 2147483647) , p("A" | Irule1020 | 2147483647) , PRest } ,
    transitions{ t("T" | AidT1 ) : TRest } ,
    pre{ (t("T" | AidT1 ) --> p("A" | Irule1020 | 2147483647)) , MTupleRest1 } ,
    post{ (t("T" | AidT1 ) --> p("A" | Irule1017 | 2147483647)) , MTupleRest2 } ,
    marking{ p("A" | Irule1017 | 2147483647) ; MRest }
    )
    rule(
    l( net(
    places{ p("A" | Irule2017 | 2147483647) , p("A" | Irule2020 | 2147483647) } ,
    transitions{ t("T" | Irule2024) } ,
    pre{ (t("T" | Irule2024) --> p("A" | Irule2017 | 2147483647)) } ,
    post{ (t("T" | Irule2024) --> p("A" | Irule2020 | 2147483647)) } ,
    marking{ p("A" | Irule2017 | 2147483647) }
    ) ) ,
    r( net(
    places{ p("A" | Irule3017 | 2147483647) , p("A" | Irule3020 | 2147483647) } ,
    transitions{ t("T" | Irule3026) } ,
    pre{ (t("T" | Irule3026) --> p("A" | Irule3020 | 2147483647)) } ,
    post{ (t("T" | Irule3026) --> p("A" | Irule3017 | 2147483647)) } ,
    marking{ p("A" | Irule3017 | 2147483647) }
    ) ) )
    | RRest
    NewMaxID
    StepSize
    aidPlace{ AidPRest }
    aidTransition{ AidTRest2 }
    if        AidTRest1 := addOldID(AidTRest | Irule1024) /\
       AidT1 := getAid(AidTRest1 | MaxID | StepSize) /\
       AidTRest2 := removeFirstElement(AidTRest1 | MaxID | StepSize) /\       NewMaxID := correctMaxID(MaxID | StepSize | 2) .

endm
\end{lstlisting}
\newpage

\section*{Reconfigurable Petri Nets - PROP.maude}

\begin{lstlisting}[language=maude]
mod PROP is
  including RULES .

  including SATISFACTION .

  subsort Configuration < State .

  var I MaxID StepSize : Int .
  var Pre : Pre .
  var Post : Post .
  vars P P1 P2 PRest PreValue : Places .
  vars T T1 T2 TRest : Transitions .
  vars MappingTuple MTupleRest1 MTupleRest2 : MappingTuple .
  vars M MRest : Markings .
  var Rules : Rule .
  var aidP : IDPoolPlace .
  var aidT : IDPoolTransition .

  op reachable : Markings -> Prop .
  
  eq net(P ,
         T ,
         Pre ,
         Post ,
         marking{ M ; MRest } ) 
     Rules
     MaxID
     StepSize
     aidP
     aidT
     |= reachable(M) = true .

  op t-enabled : -> Prop .
  
  eq net(P ,
         T ,
         pre{ (T1 --> PreValue) , MappingTuple } ,
         Post ,
         marking{ PreValue ; MRest } ) 
     Rules
     MaxID
     StepSize
     aidP
     aidT
     |= t-enabled = true .
  eq C |= t-enabled = false [owise] . 

  op enabled : -> Prop .
  
  eq net(P ,
         T ,
         pre{ (T1 --> PreValue) , MappingTuple } ,
         Post ,
         marking{ PreValue ; MRest } ) 
     Rules
     MaxID
     StepSize
     aidP
     aidT
     |= enabled = true .

  vars Irule2017 Irule2020 Irule2024 : Int .

  eq net(places{ p("A" | Irule2017 | 2147483647) , p("A" | Irule2020 | 2147483647) , P } ,
         transitions{ t("T" | Irule2024) : T } ,
         pre{ (t("T" | Irule2024) --> p("A" | Irule2017 | 2147483647)) , MTupleRest1 } ,
         post{ (t("T" | Irule2024) --> p("A" | Irule2020 | 2147483647)) , MTupleRest2 } ,
         marking{ p("A" | Irule2017 | 2147483647) ; M } )
         Rules
         MaxID
         StepSize
         aidP
         aidT
         |= enabled = true .

  var C : Configuration .
  var Prop : Prop .
  eq C |= Prop = false [owise] .

endm
\end{lstlisting}
\newpage

\section*{Reconfigurable Petri Nets - NET.maude}

\begin{lstlisting}[language=maude]
mod NET is
  including PROP .
  including LTLR-MODEL-CHECKER .

  ops initial : -> Configuration .

  eq initial =
  net(
    places{ p("A" | 3 | 2147483647) , p("A" | 4 | 2147483647) , p("A" | 2 | 2147483647) } ,
    transitions{ t("T" | 7) : t("T" | 5) : t("T" | 6) } ,
    pre{ (t("T" | 7) --> p("A" | 3 | 2147483647)) , (t("T" | 5) --> p("A" | 4 | 2147483647)) , (t("T" | 6) --> p("A" | 2 | 2147483647)) } ,
    post{ (t("T" | 7) --> p("A" | 2 | 2147483647)) , (t("T" | 5) --> p("A" | 3 | 2147483647)) , (t("T" | 6) --> p("A" | 4 | 2147483647)) } ,
    marking{ p("A" | 3 | 2147483647) ; p("A" | 4 | 2147483647) }
    )
  rule(
    l( net(
    places{ p("A" | 17 | 2147483647) , p("A" | 20 | 2147483647) } ,
    transitions{ t("T" | 24) } ,
    pre{ (t("T" | 24) --> p("A" | 17 | 2147483647)) } ,
    post{ (t("T" | 24) --> p("A" | 20 | 2147483647)) } ,
    marking{ p("A" | 17 | 2147483647) }
    ) ) ,
    r( net(
    places{ p("A" | 17 | 2147483647) , p("A" | 20 | 2147483647) } ,
    transitions{ t("T" | 26) } ,
    pre{ (t("T" | 26) --> p("A" | 20 | 2147483647)) } ,
    post{ (t("T" | 26) --> p("A" | 17 | 2147483647)) } ,
    marking{ p("A" | 17 | 2147483647) }
    ) ) )
    26
    10
    aidPlace{ (26 , (27 , (28 , (29 , (30 , (31 , (32 , (33 , (34 , (35 , (36))))))))))) }
    aidTransition{ (26 , (27 , (28 , (29 , (30 , (31 , (32 , (33 , (34 , (35 , (36))))))))))) }    .

endm
\end{lstlisting}

\end{appendices}

\end{document}